\documentclass[12pt,preprint]{emulateapj}
\usepackage{graphicx}
\usepackage{subfigure} 
\usepackage{natbib} 
\usepackage{float}
\usepackage{apjfonts}

\shorttitle{Crab Nebula's Northern Jet} 
\shortauthors{Black and Fesen}

\begin{document}

\title{A 3D Kinematic Study of the Northern Ejecta ``Jet'' of
the Crab Nebula}

\author{Christine S.\ Black \& Robert A.\ Fesen } 
\affil{ 6127 Wilder Lab,
Department of Physics \& Astronomy, Dartmouth College, Hanover, NH 03755 }

\begin{abstract}

We present moderate resolution [\ion{O}{3}] $\lambda\lambda$4959,5007 line
emission spectra of the Crab Nebula's northern ejecta jet. These data along
with an [\ion{O}{3}] image of the Crab Nebula were used to build 3-dimensional
kinematic maps of the jet and adjacent remnant nebulosity to better understand
the jet's properties and thus its likely origin.  We find the jet's systemic
velocity to be $+170$ $\pm$ 15 km s$^{-1}$ with radial velocities ranging from
$-190$ to +480 km s$^{-1}$.  Our data indicate that the jet consists of thin
filamentary walls ($V_{exp}$ $\simeq$ 40 -- 75 km s$^{-1}$), is virtually
hollow in [\ion{O}{3}] emission, and elliptical and funnel-like in shape rather
than a straight cylindrical tube as previously thought.  Examination of the
Crab's 3D filamentary structure along the jet's base reveals a large and nearly
emission-free opening in the remnant's thick outer ejecta shell.  The jet's
blueshifted and redshifted sides are surprisingly well defined and, like the
jet's sharp western limb, appear radially aligned with the remnant's center of
expansion. These alignments, along with the opening in the nebula at the jet's
base and proper motions indicating an expansion age in line with the 1054
supernova event, suggest a direct connection between the jet's formation and
the Crab's radial expansion.  While our analysis supports the scenario that the
jet may simply represent the highest velocity material of the remnant's N-S
bipolar expansion, the nature of this expansion asymmetry remains unclear.

\end{abstract}


\section{Introduction}

The Crab Nebula was the first supernova remnant to be associated with a ``guest
star'' seen by ancient Chinese and Japanese astronomers in 1054 AD
\citep{hub28,lun38,duy42}, the first pulsar firmly associated with a
supernova remnant \citep{com69,CocDisTay69} as well as the first astronomical
object found to emit synchrotron radiation \citep{shk53,shk57,dom54}. Despite
being one of the best studied astronomical objects, a few properties
of the Crab have not yet been fully understood (see reviews by \citealt{DavFes85}, \citealt{hes08}, and \citealt{BuhBla14}). 
 
One of these is the nature of a curious filamentary feature 45$''$ wide and
extending approximately 100$''$ off the nebula's northern limb. First reported
as a ``faint jetlike structure'' seen in deep optical images by
\citet{bergh70}, narrow passband optical images showed it consists of line
emitting filaments which are brightest in [\ion{O}{3}] $\lambda\lambda$4959,
5007 \citep{CheGul75,GulFes82}. Subsequent deep radio and optical continuum
studies showed the presence of associated extremely faint synchrotron emission
\citep{vel84,WolVC87}.

Follow-up spectral and kinematic studies show the feature to be 
hollow \citep{FesGul83,shu84}, with a mean radial velocity expansion of 260 km
s$^{-1}$, and a mean heliocentric velocity of 184 km s$^{-1}$ \citep{Mar90}. It
also appears to be slightly tilted into the plane of the sky by eight
degrees \citep{shu84} and, although large in extent, has a relatively small
estimated mass $\approx$ 0.003 M$_{\odot}$ \citep{VCVerWol85}.

A particularly interesting but puzzling aspect of this feature is its parallel
edge morphology making it appear more like a cylinder rather than a diverting
outflow of radially expanding ejecta. While its sharp western edge closely
aligns back to the Crab's estimated center of expansion
\citep{FesGul86,FesSta93}, its central axis is unaligned with either the center
of expansion or the remnant's pulsar.  Although sometimes
referred to as a `trail', `spur', `stem', or `chimney'
\citep{bla83,kun83,morRob85}, here we will use the word `jet' as it was
originally described by \citet{bergh70}, but this should not be taken as
necessarily implying a narrow or directed beam of material or energy.

A variety of explanations have been proposed to explain the nature of the
Crab's jet (see Table \ref{tab:thry}).  One early theory involved a
two-component pulsar wind nebula (PWN) model where dense, line radiation
emitting filaments were frozen in a ``strong'' 10$^{-3}$ Gauss magnetic field.
Field lines forced between the filaments via plasma instabilities then expand
outward forming the jet \citep{byc75,Mar90}.  Other models suggest a
combination of magnetic fields and the Crab's PWN leading to a
Rayleigh-Taylor instability `bubble' bursting through the nebula's shell of
ejecta \citep{CheGul75,SanHes97}.  

While some theories invoke instabilities within the Crab's structure to
describe the formation and uniqueness of the jet, alternative hypotheses
suggested that the ambient interstellar medium (ISM) surrounding the nebula
contributed to the jet's formation.  In these models, the jet's formation was
attributed to Rayleigh-Taylor instabilities in the remnant's shell of filaments
that result as the remnant expanded into a low density region in the local ISM
\citep{kun83,VCVerWol85}.

%Theory Table
%\begin{center}
\begin{deluxetable*}{ll}[t]
  \tablecolumns{2}
  \tablecaption{Proposed Explanations for the Jet's Formation\label{tab:thry}}
  \tablehead{
  \colhead{Theory} & \colhead{References}\\}
  \startdata
        Filaments in Expanding Magnetic Field & \citet{byc75,Mar90}\\
        PWN Driven Instability and Breakout       & \citet{CheGul75,SanHes97,Smi13}\\
        Mass-loss Trail from Red Giant Progenitor & \citet{bla83}\\
        Expansion into a Low Density ISM Region  & \citet{kun83,VCVerWol85}\\
        Relativistic Pulsar Beam & \citet{shu84,ben84}  \\
                                 & \citet{mic85,biekro90} \\
        Interaction with a Local Interstellar Cloud & \citet{morRob85} \\
        Highest-Velocity Ejecta of N-S Bipolar Expansion & 
\citet{FesSta93,FesShuHur97,RudFesYam08}
   \enddata
\end{deluxetable*}
%\end{center}

%Figure 1 SLITS
\begin{figure*}[t]
        \centering
        \subfigure{
                \includegraphics[width=2\columnwidth]
                {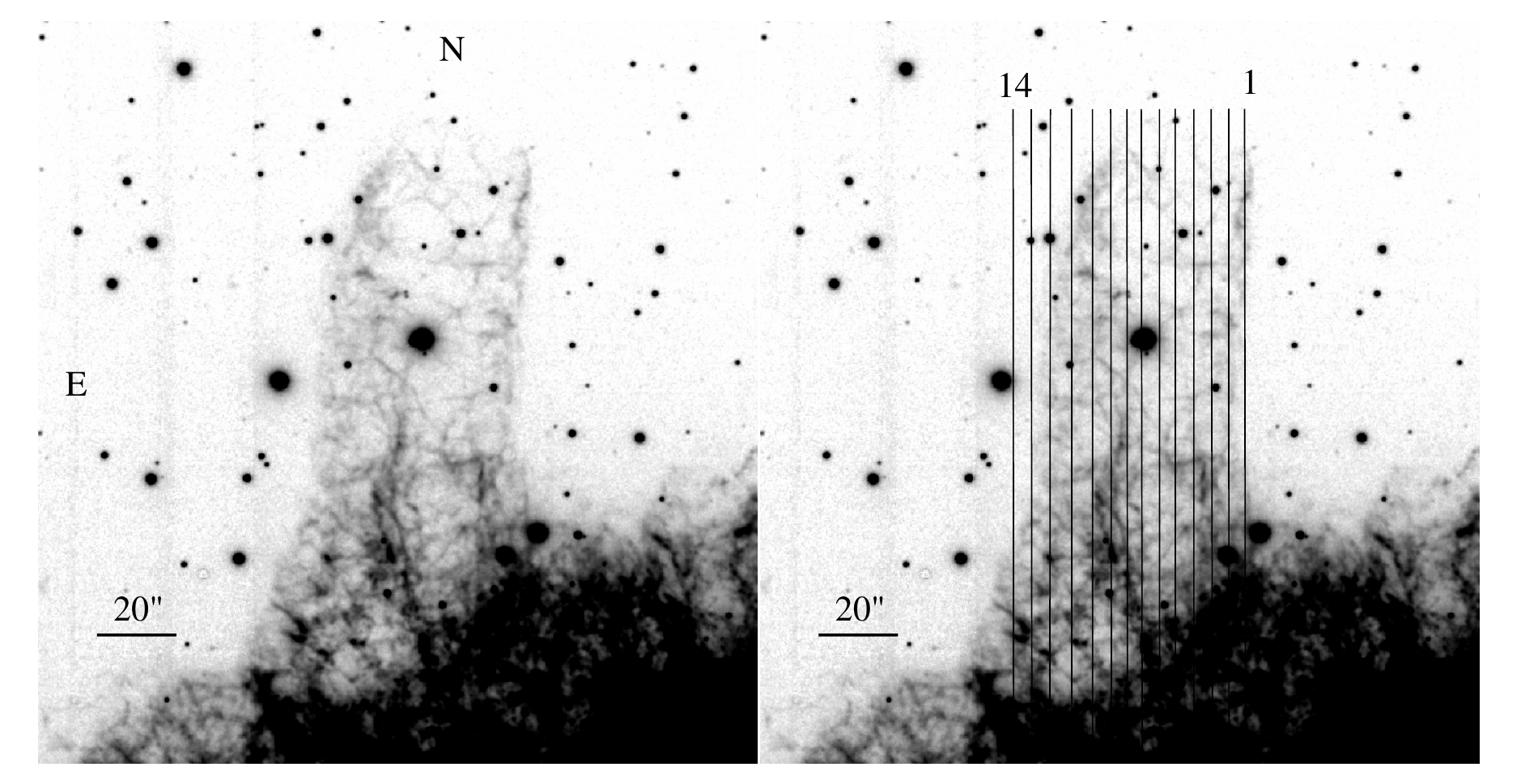} }
        \hfill
        \subfigure{
                \includegraphics[trim=0.15cm 0cm 0.15cm 1cm,width=2.\columnwidth]
                {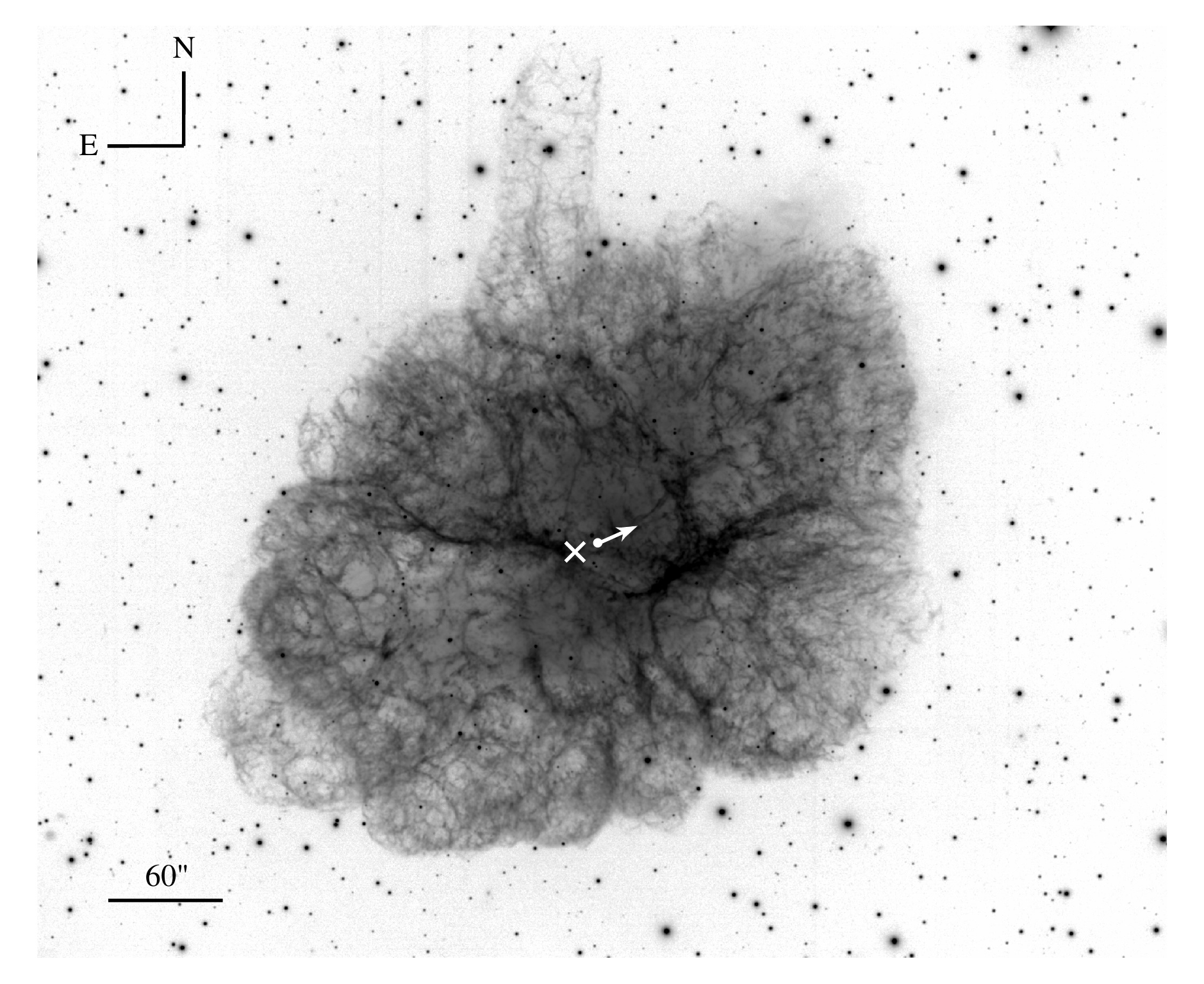} }
        \caption{A 2005 [\ion{O}{3}] $\lambda$5007 image of the Crab Nebula's northern
        ejecta ``jet" (upper left panel) from \citet{RudFesYam08} with the 14 north-south
        slit positions overlaid (upper right panel). Bottom panel: Image of the whole Crab Nebula showing 
the northern jet with respect to the entire remnant. The `X' marks the \citet{Nug98} center of expansion and the 
white dot and arrow marks the pulsar and its proper motion direction as determined by \citet{Kap08}.}
        \label{fig:slits}
\end{figure*}

Other proposals include a relativistic plasma beam resulting from magnetic
fields directly off the pulsar \citep{shu84,ben84,mic85,biekro90} or that the
jet is just the highest velocity component of the Crab's N-S bipolar expansion
\citep{FesSta93,RudFesYam08}.  A mass-loss trail from the red giant progenitor
star was proposed by \citet{bla83}, and \citet{morRob85} suggested that it was
formed as a result of the interaction between a small local interstellar cloud
creating a `shadowed flow'.

In order to explore its kinematic properties as a means of better constraining
its likely origin, we obtained moderate dispersion, long slit spectra of its
[\ion{O}{3}] line emission.  These data allow us to investigate in detail the
jet's kinematic structure in relation to the main nebula along the Crab's
northern limb and whether the jet's structure is connected to the Crab's
expansion point.  Our observations are described in
\textsection{\ref{sec:Obs}}, our results and 3-dimensional reconstructions
presented in \textsection{\ref{sec:Analysis}}, our analysis in
\textsection{\ref{sec:Res}}, our discussion of the results in
\textsection{\ref{sec:Disc}}, and our conclusions in
\textsection{\ref{sec:Conc}}.

\section{Observations} \label{sec:Obs}

Optical spectra of the Crab's northern jet feature were taken in October 2013
with the 2.4m Hiltner telescope at the MDM Observatory on Kitt Peak, AZ  using
a Boller and Chivens CCD Spectrograph (CCDS) and a 1800 grooves mm$^{-1}$ grating
blazed at 4700 \AA \ yielding 0.275 \AA \ per pixel. 
A $1\farcs 2$ x 264$''$ N-S slit was placed along the
jet's western edge with the bottom of the slit approximately 50$''$ above the
center of expansion as defined by \citet{Nug98} ([J2000] $05^{\rm h} 34^{\rm m}
32.84^{\rm s}$ +22$^{\circ}$ 00$'$ $48\farcs0$).  

Fourteen 900s spectra were taken covering the width of the jet, moved
approximately 5$''$ between each exposure.  The slit positions are shown in
Figure \ref{fig:slits}.  The resulting spectra have a FWHM resolution of 0.7
\AA\ ($\simeq$ 40 km s$^{-1}$) and covered 330 \AA\ which included H$\beta$ and
[\ion{O}{3}] $\lambda\lambda $4959, 5007 and a scale of $0\farcs 363$
pixel$^{-1}$. Velocity measurements are believed accurate to $\pm 15$ km
s$^{-1}$ for all but the faintest jet emission features. Comparison spectra for
wavelength calibration were  taken using the Ar, Hg, Ne, and Xe lamps.  Data
reduction was done using IRAF\footnote{IRAF is distributed by the National
Optical Astronomy Observatories, which are operated by the Association of
Universities for Research in Astronomy, Inc., under cooperative agreement with
the National Science Foundation.} and consisted of bias and background
subtraction, wavelength calibration, and cosmic ray removal.

\altaffiltext{2}{IRAF is distributed by the National Optical Astronomy Observatories, 
    which are operated by the Association of Universities for Research 
    in Astronomy, Inc., under cooperative agreement with the National 
    Science Foundation.}

%Figure Slit45
\begin{figure*}
	\centering
	\includegraphics[width=0.75\textwidth]{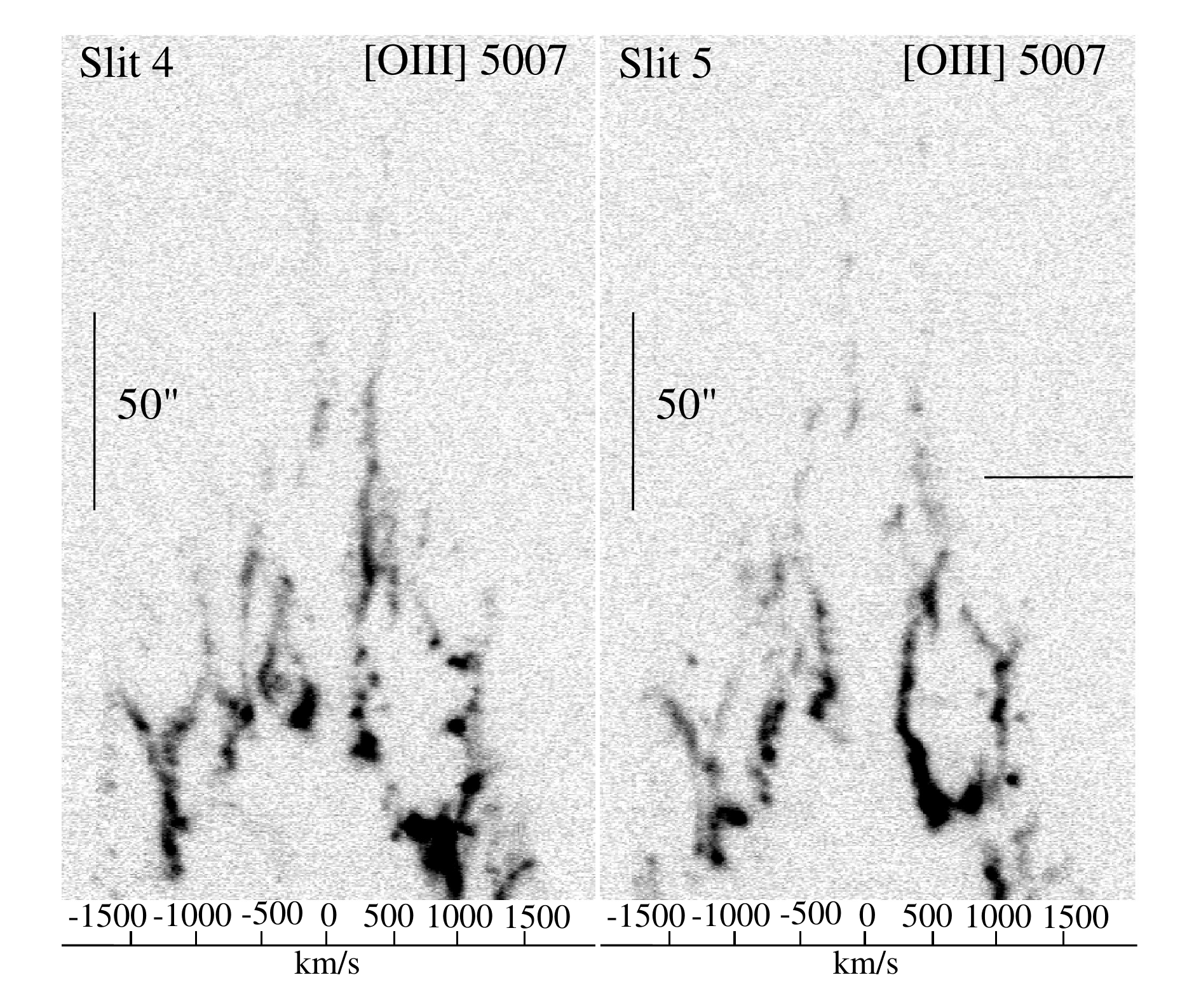} 
	\caption{Spectra for slit positions 4 and 5 showing the $\lambda$4959 
	subtracted $\lambda$5007 [\ion{O}{3}] line emission structure.  
In the right panel, the horizontal line marks the base of the jet.} 
	\label{fig:slit45} 
\end{figure*}

\section{Analysis} \label{sec:Analysis}

A 3-dimensional kinematic representation of the jet was constructed using the
jet's [\ion{O}{3}] $\lambda\lambda$4959,5007 emission lines.  Due to the nebula's
expansion velocity reaching as high as 1800 km s$^{-1}$ \citep{LunTzi12}, a portion of the most
blueshifted $\lambda$5007 emissions overlapped with the most redshifted
$\lambda$4959 emissions.  This $\lambda$4959 emission line overlap was removed by
assuming a I(5007)/I(4959) ratio of 2.91 \citep{Ost06}.  Figure
\ref{fig:slit45} shows an example of the reduced data.

%Figure Ring Positions
\begin{figure}[b]
        \centering 
        \includegraphics[width=0.5\textwidth]{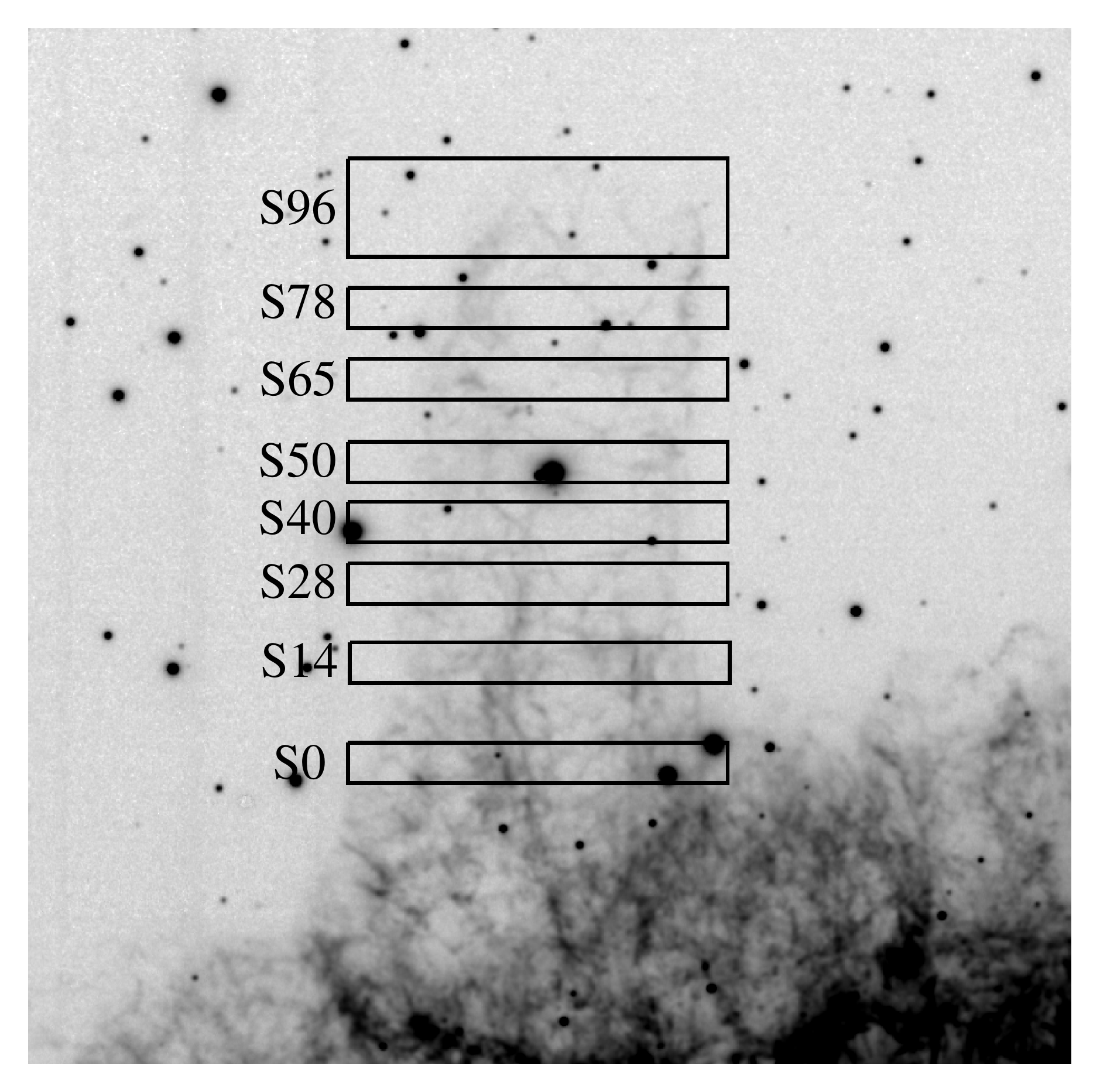}
        \caption{Positions of our selected jet sections.}
        \label{fig:sections}
\end{figure}

%Figure Each Ring
\begin{figure*}[t]
	\centering 
	\subfigure[]{ \label{fig:bottom}
		\includegraphics[width=.85\columnwidth]
		{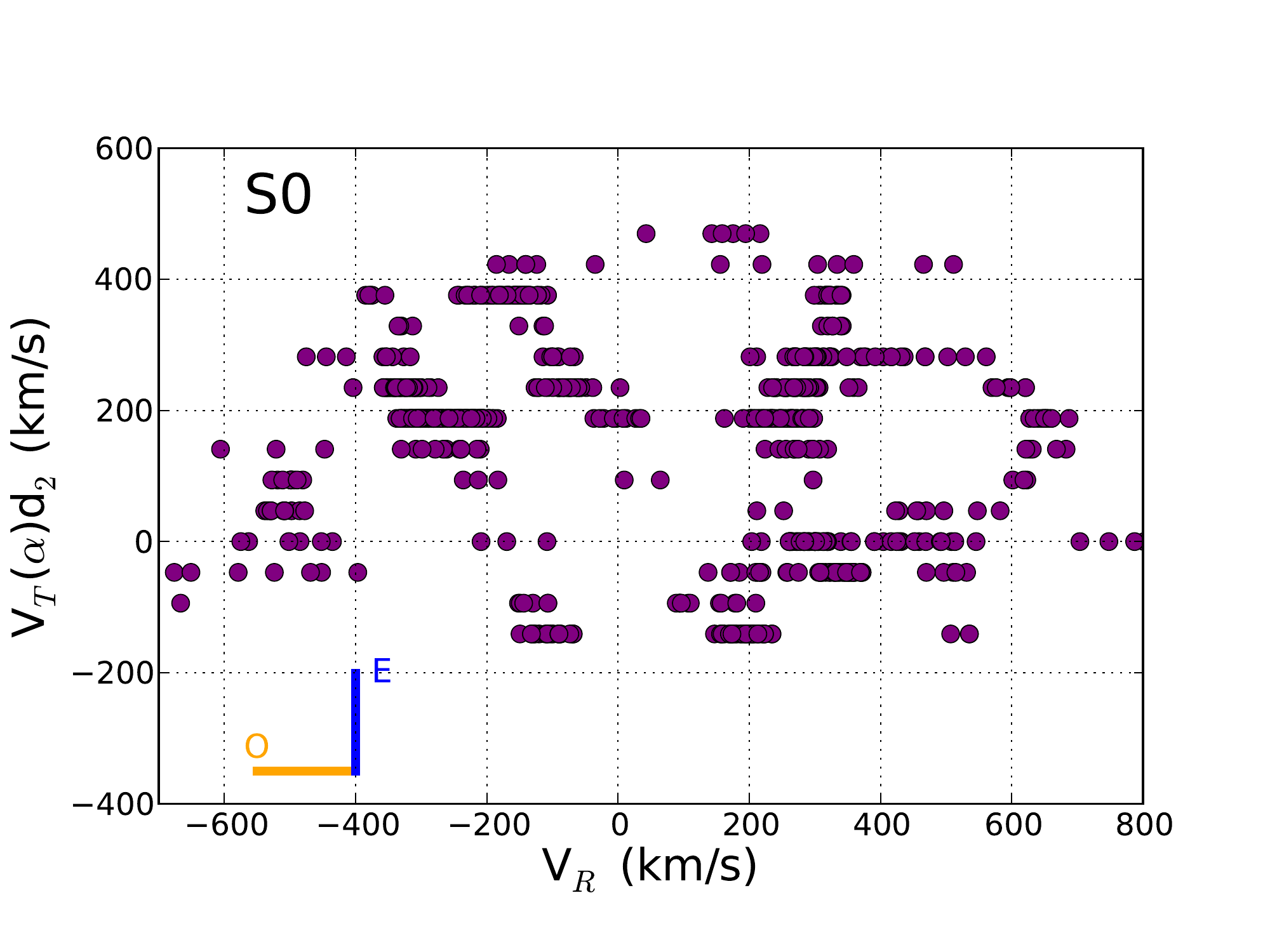} }
	\subfigure[]{ \label{fig:violet} 
		\includegraphics[width=.85\columnwidth]
		{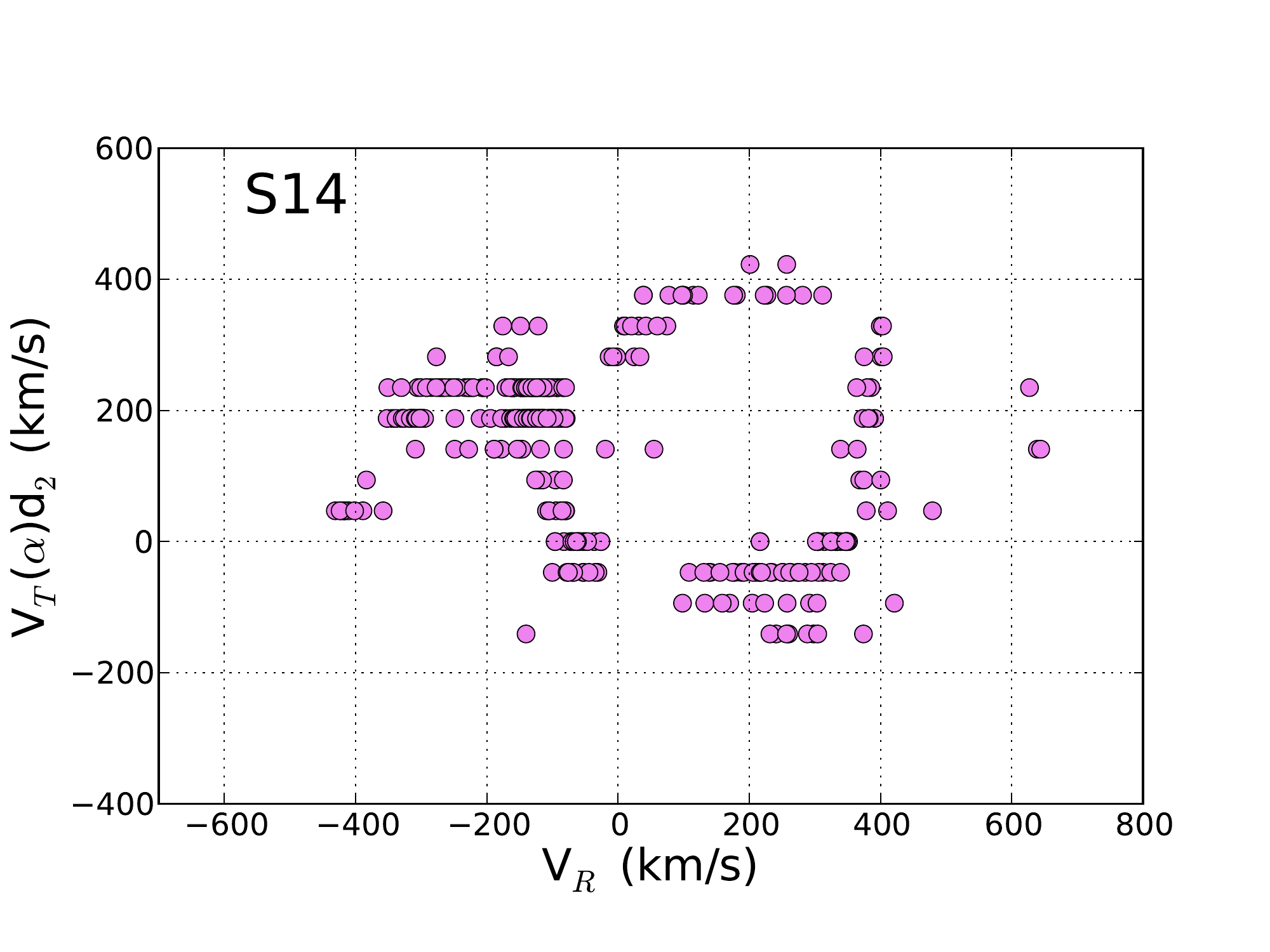} } 
	\subfigure[]{ \label{fig:blue} 
		\includegraphics[width=.85\columnwidth]
		{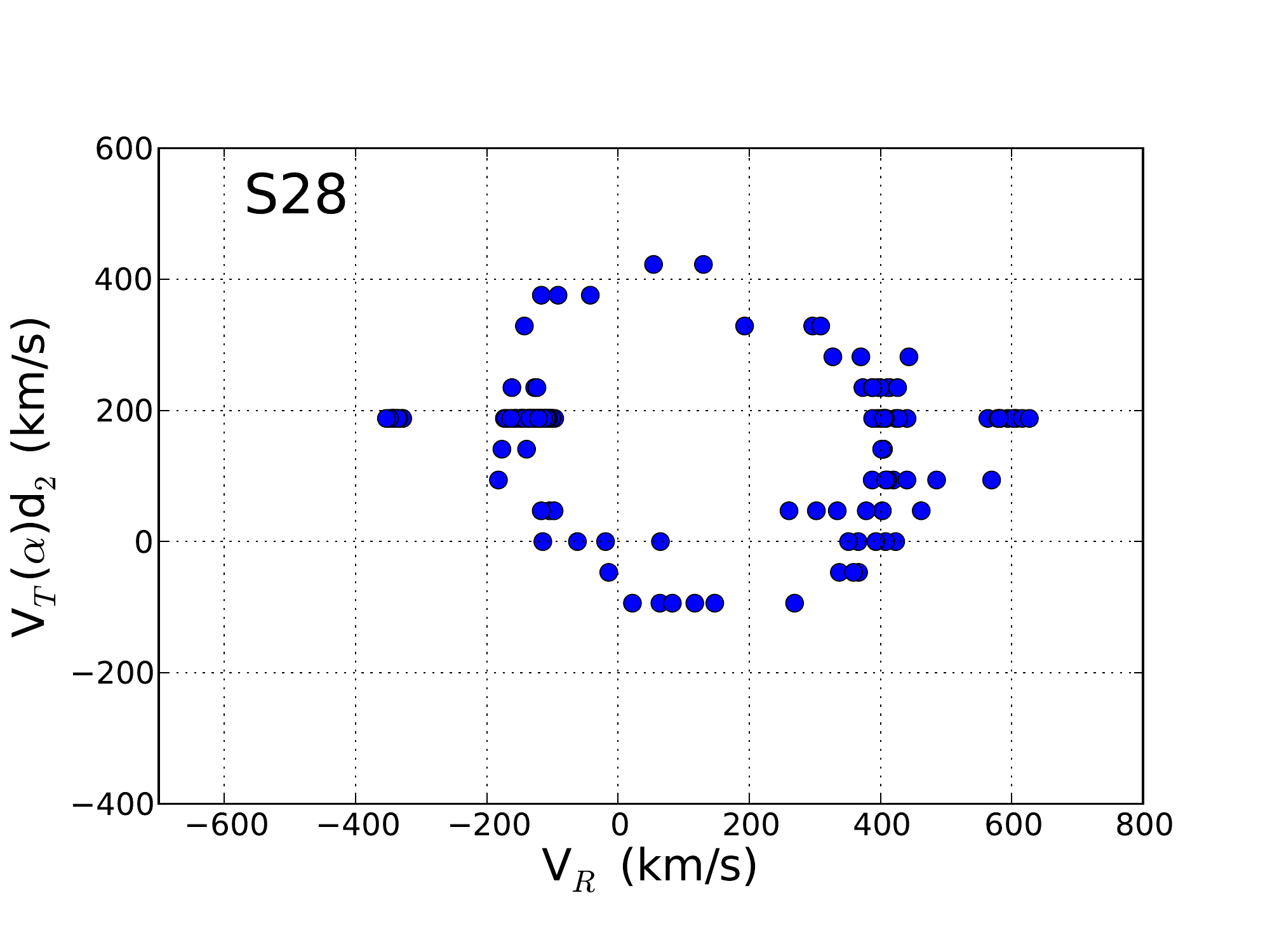} } 
	\subfigure[]{ \label{fig:cyan}
		\includegraphics[width=.85\columnwidth]
		{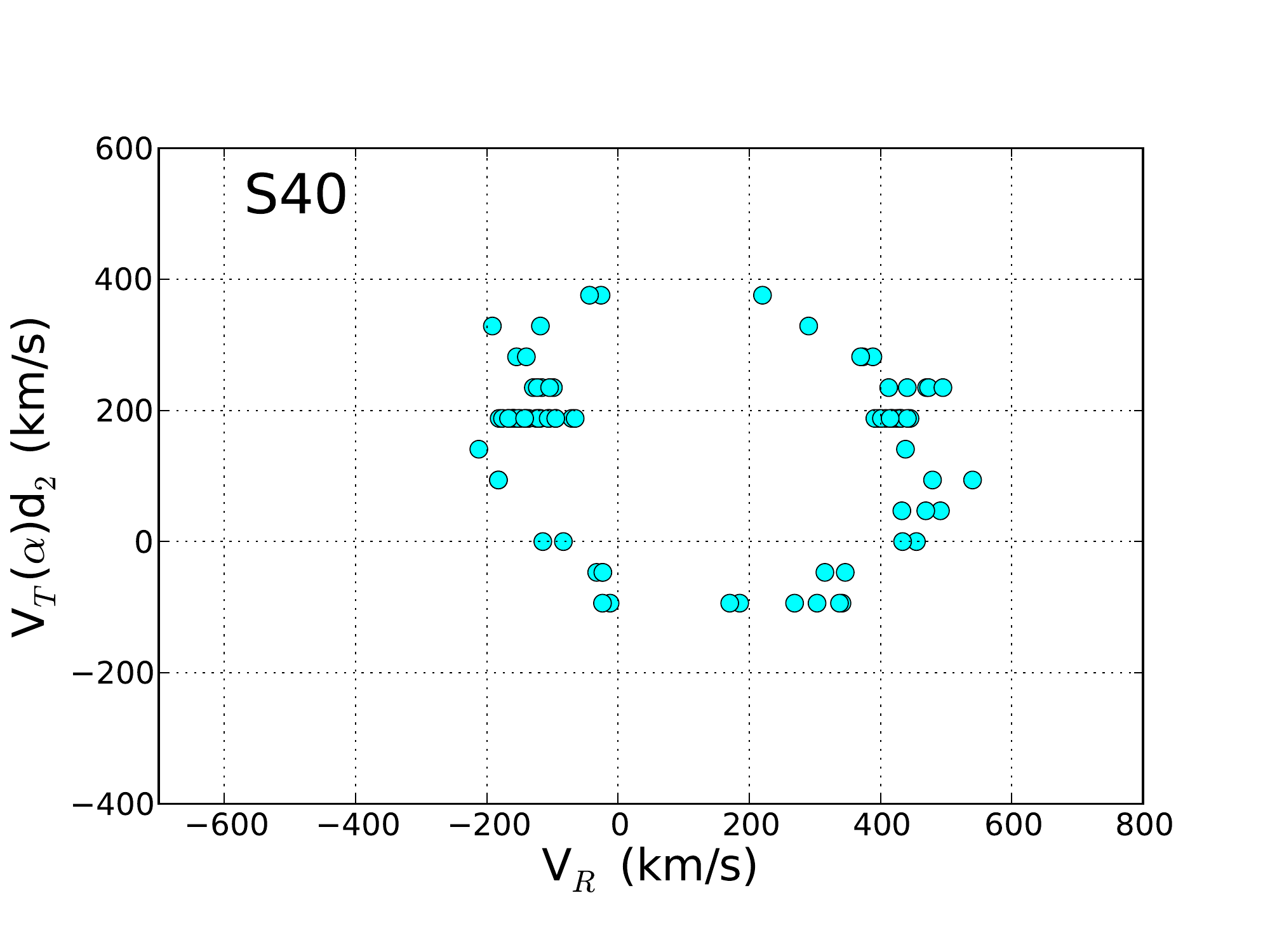} }
	\subfigure[]{ \label{fig:green}
		\includegraphics[width=.85\columnwidth]
		{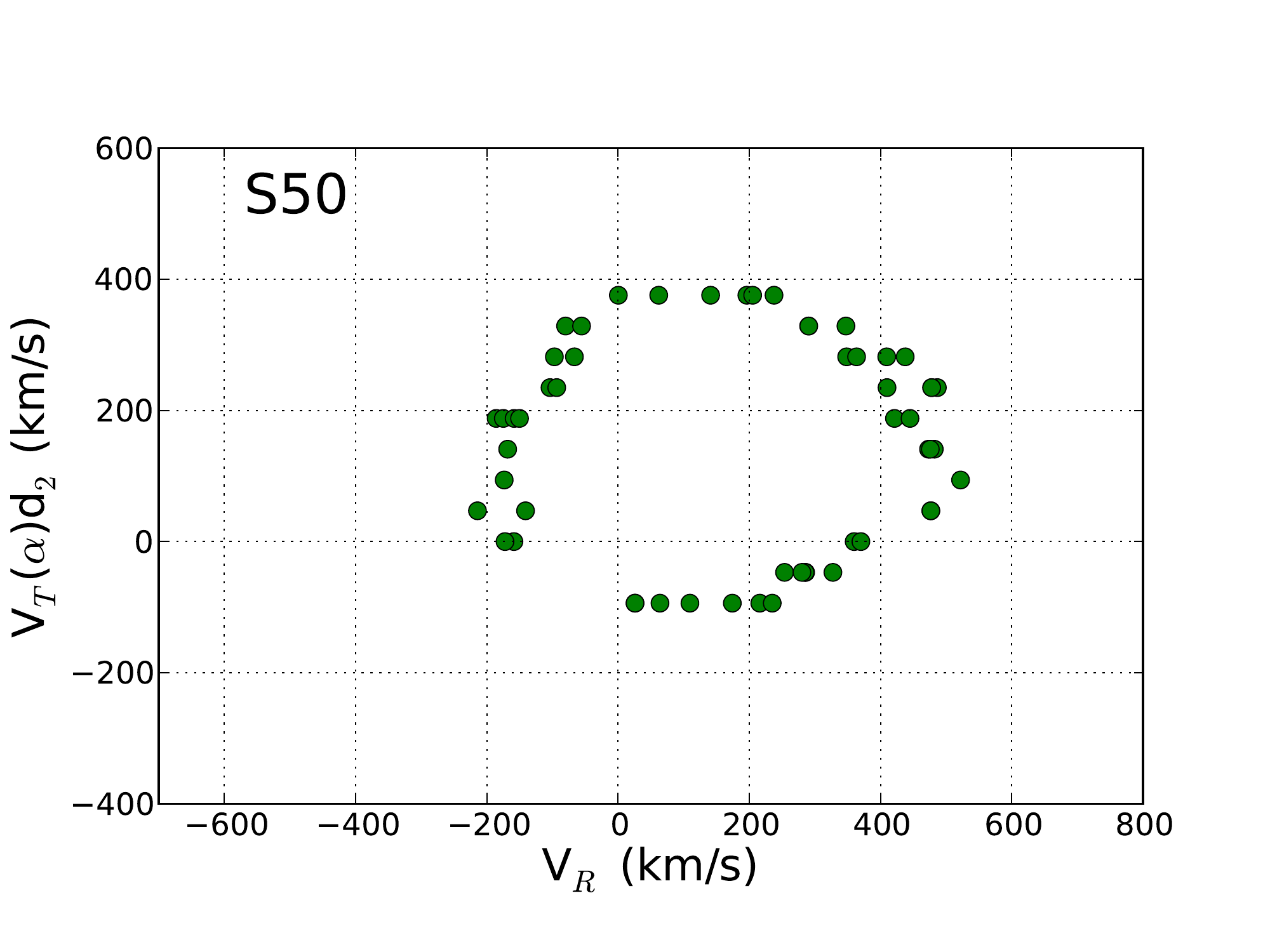} }
	\subfigure[]{ \label{fig:orange} 
		\includegraphics[width=.85\columnwidth]
		{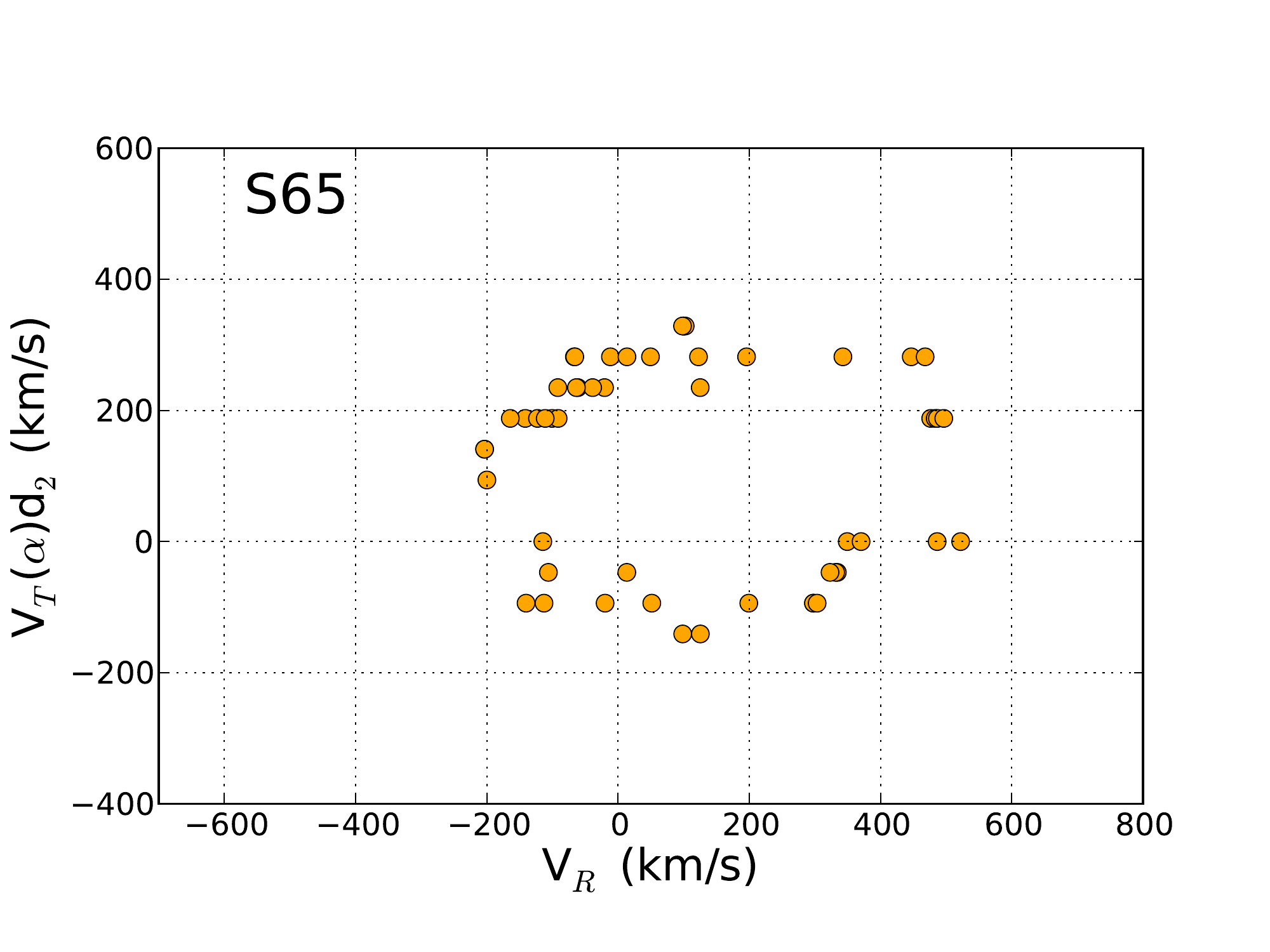} } 		 
	\subfigure[]{ \label{fig:maroon} 
		\includegraphics[width=.85\columnwidth]
		{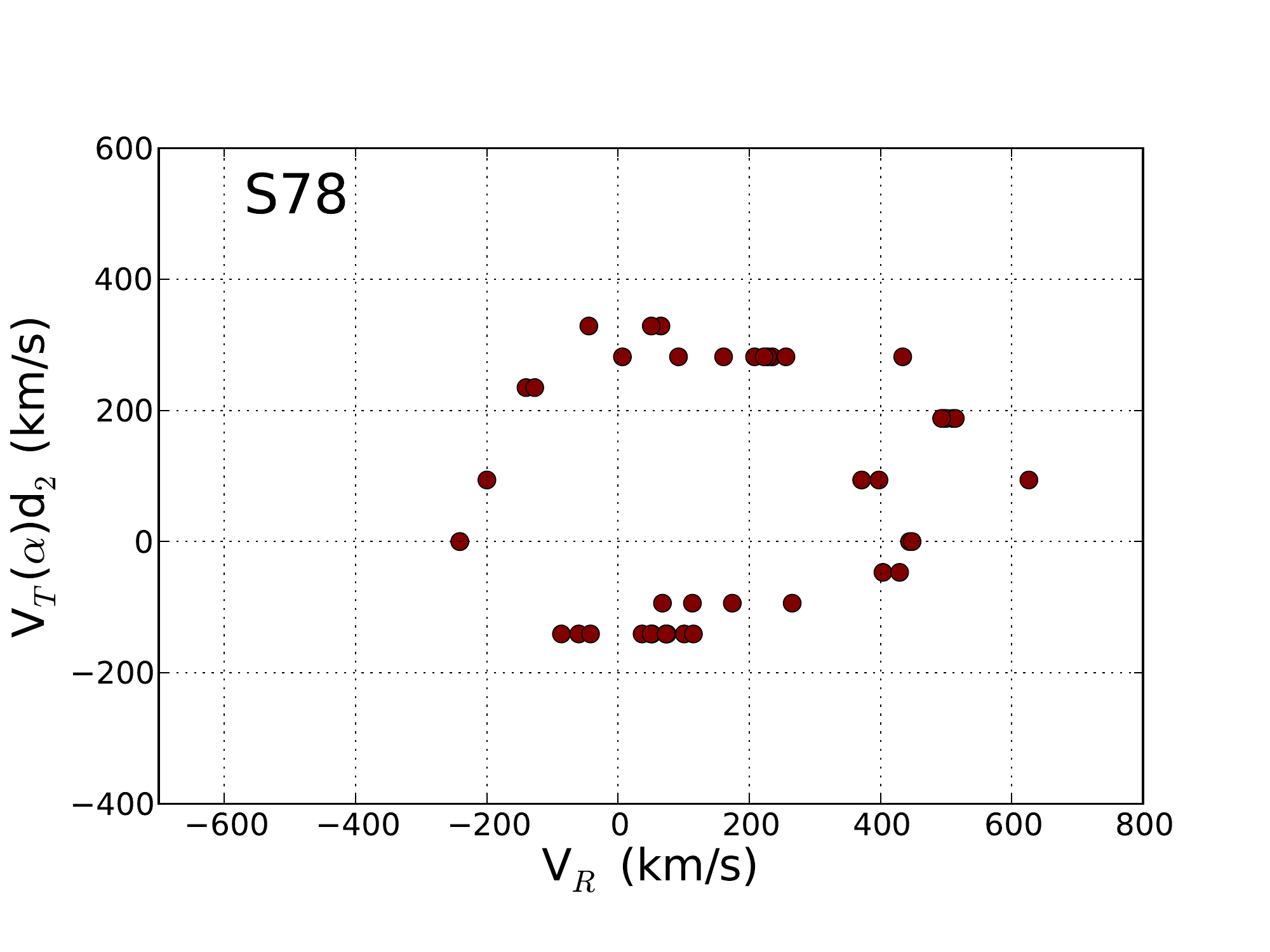} } 
	\subfigure[]{ \label{fig:red}
		\includegraphics[width=.85\columnwidth]
		{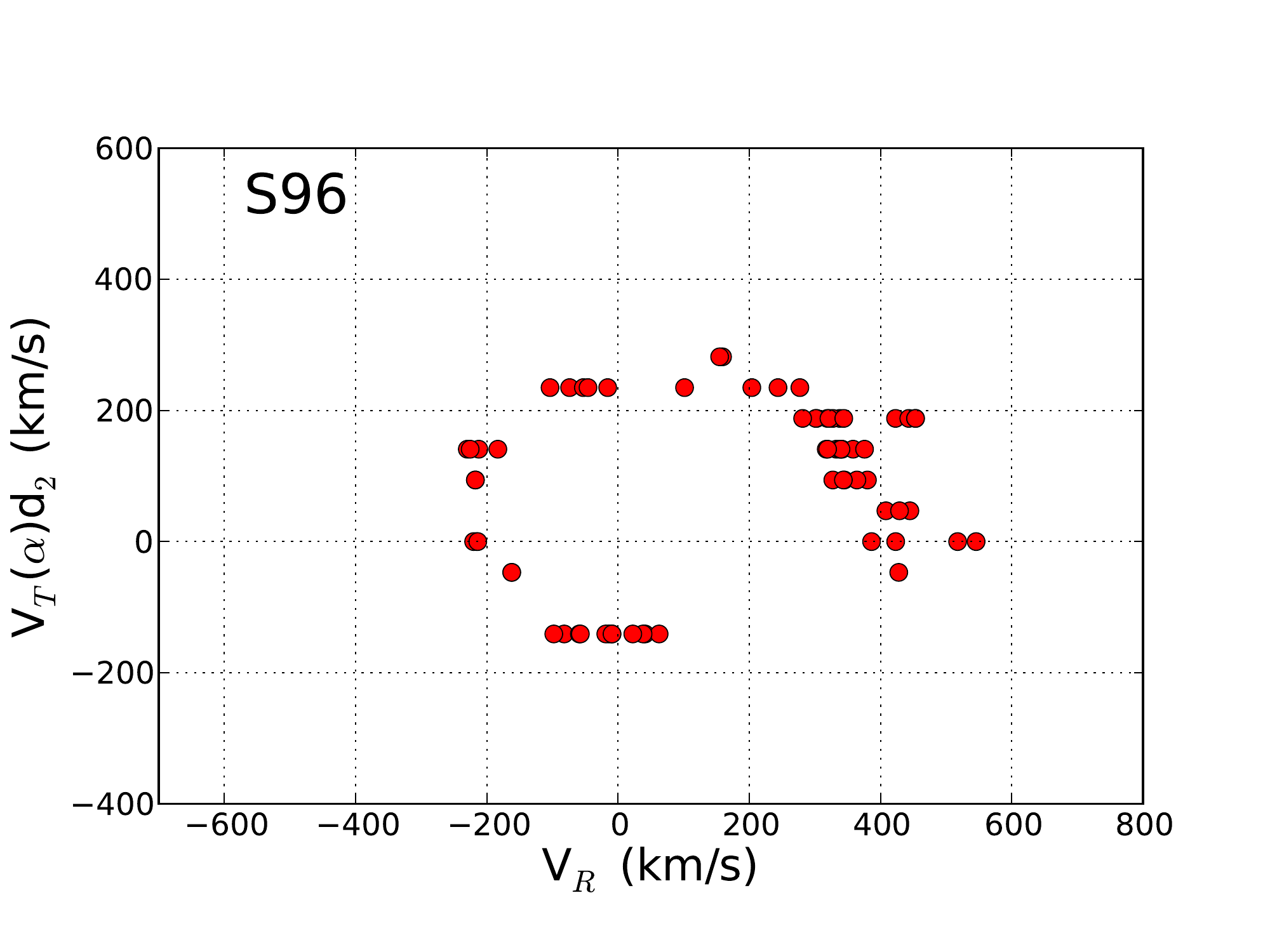} }
		\caption{Velocity plots for our eight chosen jet sections shown in Figure \ref{fig:sections} where V$_R$ is the radial velocity and V$_T(\alpha)$d$_2$ is the east$-$west transverse velocity using a distance, d, of 2 kpc. (a) S0 (b) S14 (c) S28 (d) S40 (e)S50 (f) S65 (g) S78 (h) S96.  A color version of this figure is available in the online journal. }  
		\label{fig:color_sections}
\end{figure*}

A 2005 Subaru [\ion{O}{3}] image of the Crab Nebula (Fig. 1; see
\citealt{RudFesYam08}) was used to match the jet's features with those seen in
the spectra.  This image along with an assumed distance for the Crab of
d$_2$=2.0 $\pm$ 0.5 kpc \citep{Tri68,Tri73}, a formation date of 1054 AD for the jet based on
its measured proper motions \citep{RudFesYam08}, and the center of expansion
determined by \citet{Nug98} were used to calculate the transverse velocities
for features in the jet. 

Due to its faintness, we divided the jet into two  different regions for
analysis: the jet's base comprising the interface of it with the Crab's
northern limb, and the jet feature itself.  Code developed in Python was used
to create a 3D map of the jet (from its base to the farthest extent
northward) while the Crab Nebula's northern limb and jet interface was mapped
using an IDL code and covered the region from the jet's base down well into the
Crab's interior.  For the purposes of this work, the dividing line between
these two regions was defined as the northern most star of the double stars
seen along the jet's southwestern limb ([J2000] $05^{\rm h} 34^{\rm m}
31.68^{\rm s}$ +22$^{\circ}$ 03$'$ $29\farcs61$; see Fig.  \ref{fig:slits}).

The jet's emission structure was subdivided into eight sections as shown in
Figure 3.  We use the notation SXX to indicate the distance in arcsec each
section is above the base of the jet.  Each jet section is roughly 7$''$ thick
except for S96 which is 18$''$ thick due the faintness  of features near the
jet's northern tip. 

The jet's lower and top ring sections (S0 and S96) were excluded from parts of
our kinematic analysis of the jet structure since they were either too
contaminated by the main nebula or too sparsely populated.  The remaining six
rings (S14, S28, S40, S50, S65, and S78) were used to determine the expansion
properties of the jet.

Our kinematic analysis of the jet's emission structure was developed using the
\textit{matplotlib} libraries within Python.  Data from each slit were
represented by a sample of points selected by eye along the jet's filamentary
structure.   The total sample contained over 2,700 points in an effort to
faithfully map the jet's filamentary structure as well as provide a
representation of intensity along the features.  This method has the added
benefit of removing any stellar continuum from the map.  With the construction
of this 3D map, features like the western and eastern walls could be isolated, analyzed, and viewed from various angles giving a better understanding of
the jet's overall structure.

A 3D map of [\ion{O}{3}] emission at the jet -- nebula interface was
constructed by creating a data array from the 14 slit spectra using the IDL
subroutine VOLUME.  Only pixels with intensities at least three times that of the background were used.  Dark lines running through the resulting 3D image
are the stellar continua of stars detected within the slits.

%Figure 5 Rings
\begin{figure*}
	\centering 
	\subfigure[]{ \label{fig:ring0}
		\includegraphics[trim=1.5cm 0cm 2cm .5cm,width=.9\columnwidth]
		{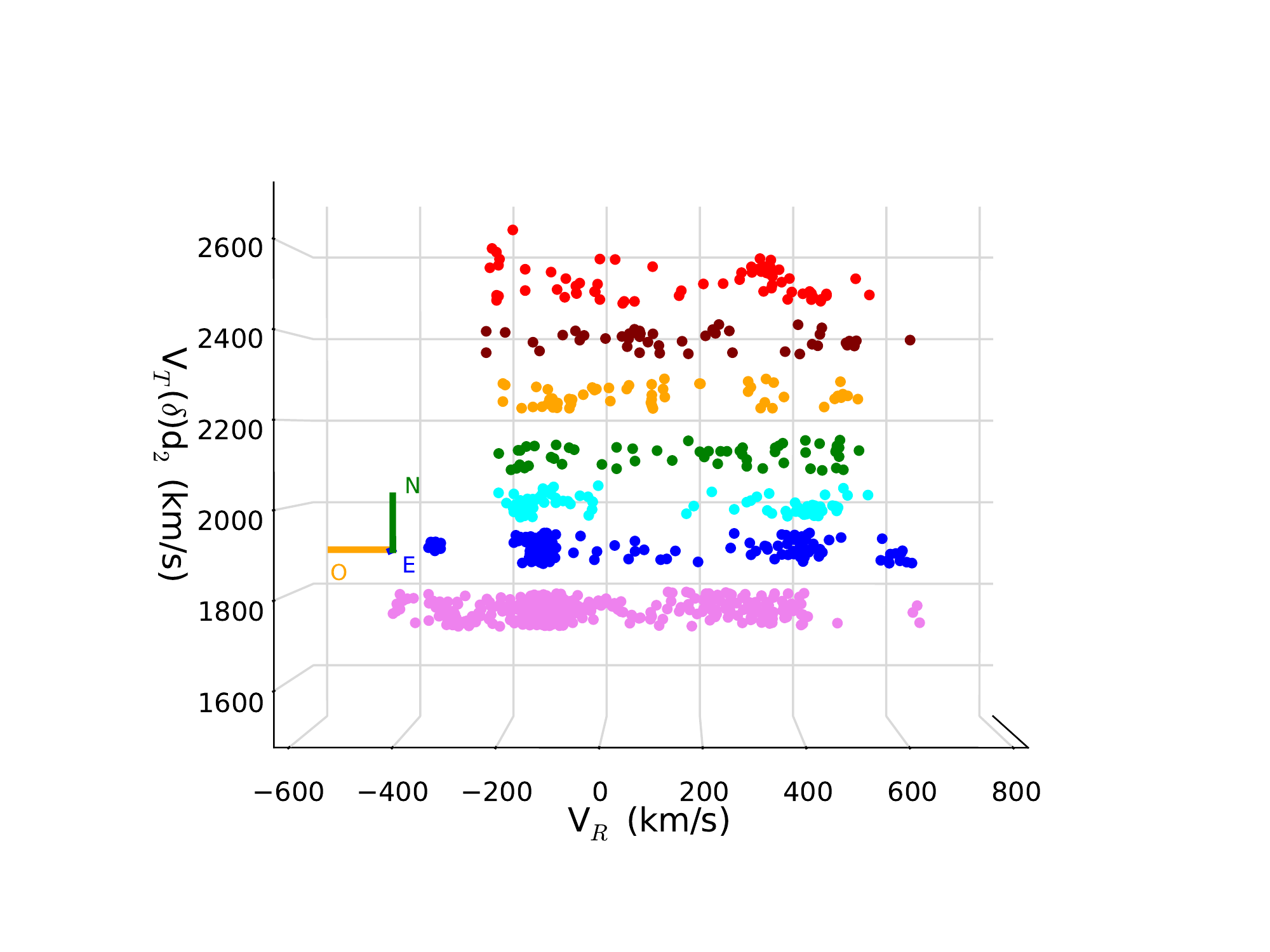} } 
	\hfill
	\subfigure[]{ \label{fig:ring45} 
		\includegraphics[trim=2cm 0cm 1.5cm .5cm,width=.9\columnwidth]
		{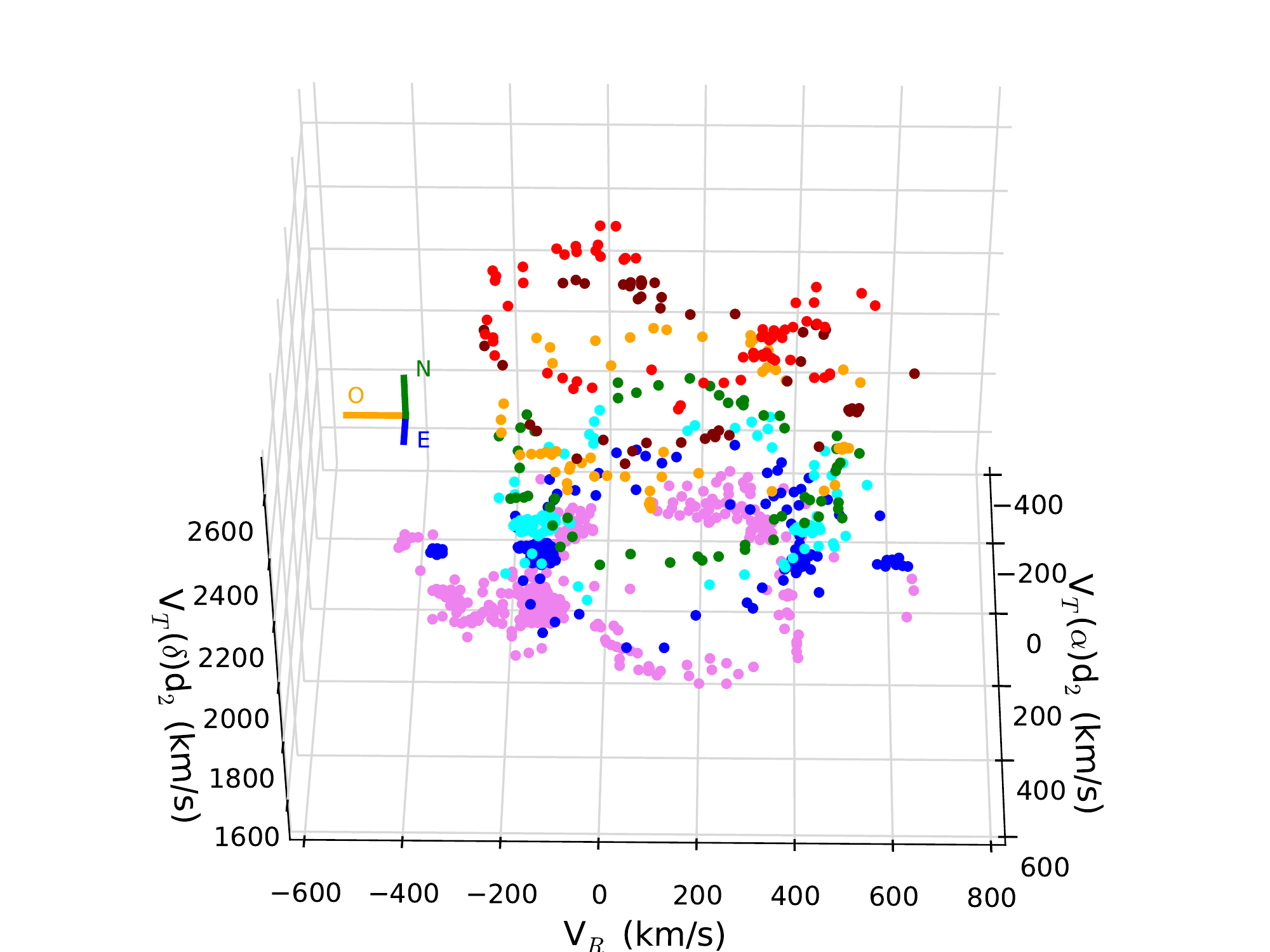} } 
	\subfigure[]{ \label{fig:ring90}
		\includegraphics[trim=1.5cm 2cm 2cm .5cm,width=.9\columnwidth]
		{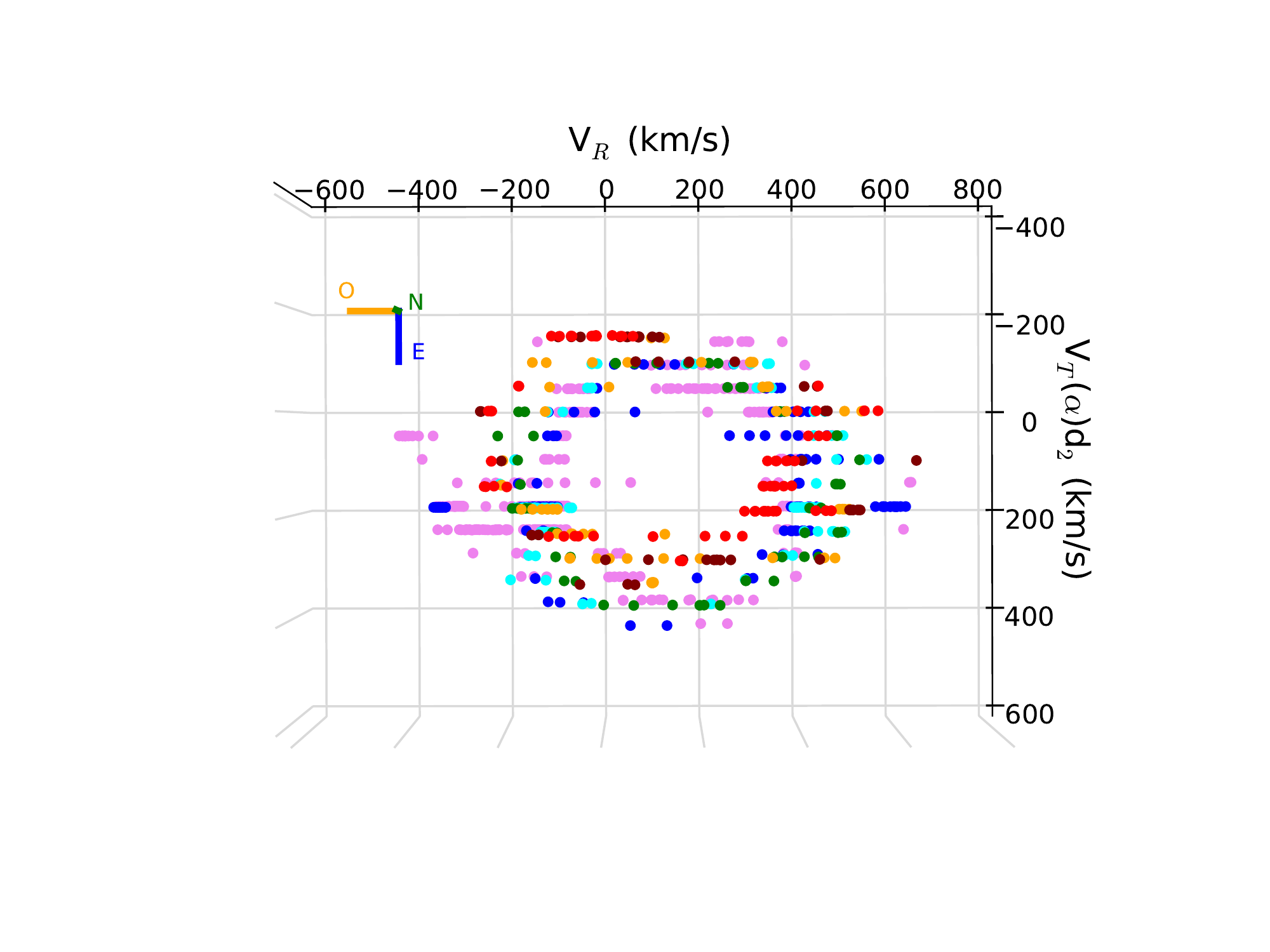} }
	\hfill
	\subfigure[]{ \label{fig:ring100} 
		\includegraphics[trim=2cm 2cm 1.5cm .5cm,width=.9\columnwidth]
		{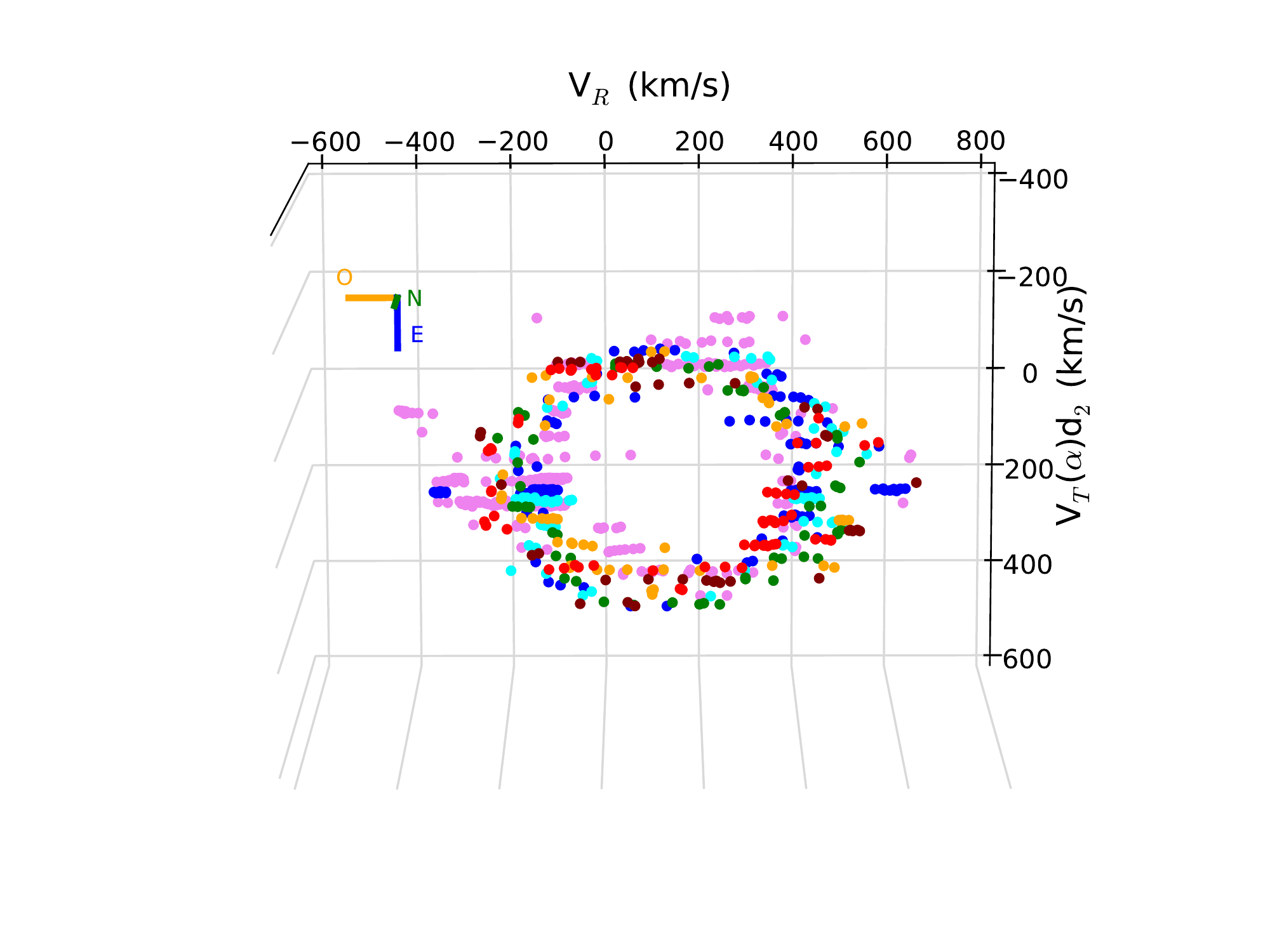} }
		\caption{Combined velocity plots of the seven upper sections (S14 in violet to S96 in red) of the Crab jet viewed at angles of: (a) 0$^{\circ}$, (b) 45$^{\circ}$, (c) 90$^{\circ}$, and (d) 100$^{\circ}$.  A color version of this figure is available in the online journal. } 
		\label{fig:rings}
\end{figure*}

\section{Results} \label{sec:Res}

\subsection{Jet Morphology}

Figure \ref{fig:color_sections} shows the eight individual jet sections plotted
in terms of radial velocity (V$_R$) and east-west transverse velocity (V$_T(\alpha)$d$_2$) relative to the
remnant's expansion point \citep{Nug98} where the sections are offset to give the portion of the jet directly above the nebula's center of expansion zero radial velocity.
The sections S0, S14, S28, S40, S50, S65, S78, and S96 are color coded
as purple, violet, blue, cyan, green, orange, maroon,
and red, respectively.

With the exception of the most southern section (S0) near the base of the jet,
these plots show the jet's structure to be remarkably empty of [\ion{O}{3}]
emission with well defined ring-like structures.  While adjacent jet sections
exhibit fairly similar appearances and dimensions, moving from just above the
jet's base (S14) to its northern tip (S96), these rings of [\ion{O}{3}]
emission become larger in diameter but also less complete.  This gradient in
diameter can easily be seen, for example, by comparing S28 (blue) to S78
(maroon) or S14 (violet) to  S96 (red).  

In contrast, the lowest jet section, S0, shows no single distinct ring of
emission, instead consists of several small cavities between filaments. This
sort of small-scale, bubble-like morphology is present throughout the remnant
so its appearance here in section S0 which includes the northern limb of the
Crab's thick ejecta shell is not unexpected. However, the apparent lack of a
hole in the S0 section coincident with the emission rings seen in the upper jet
sections is misleading since even extremely faint emissions are plotted here,
thus giving the appearance of a more filled structure.  

Figure \ref{fig:color_sections} also shows the jet's filamentary walls to be
relatively thin compared to the jet's $\simeq$650 km s$^{-1}$ radial
velocity range. We note that apparent narrowness of some ring sections in transverse
velocity is due in some measure to the limited slit coverage of our data.
Nonetheless, we find the jet walls to have a typical thicknesses of $\simeq$ 40
-- 75 km s$^{-1}$, consistent with an estimate of 60 km s$^{-1}$ reported by
\citet{Mar90} based on lower dispersion data.  However, there are several
exclusions to this range in certain ring sections where jet walls can
exceed a 100 km s$^{-1}$ range in velocity.

Fewer [\ion{O}{3}] filaments are present farther up the jet especially along its
northeastern limb. Consequently, its upper northernmost sections becomes less
complete. This incompleteness can be seen as gaps in the top three ring
sections (S65, S78, and S96) of the jet. The largest of these missing sections is
in the topmost ring (S96) which coincides with the highly curved appearance of
the jet along its northeastern tip.

Previous studies of the Crab jet found the jet to be cylindrical in shape
\citep{GulFes82,Mar90} as its near parallel limb shape might indicate.
However, our 3D mapping suggests that the jet is actually elliptical in shape
and increasingly so farther up the jet (e $\simeq$ 0.1 --  0.4) with its major
axis pointed nearly in to the plane of the sky (position angle $\simeq$
340$^{\circ}$) as can be seen in the individual sections shown in Figure
\ref{fig:color_sections}.

%Figure 6:  Python
\begin{figure*}
        \centering
        \subfigure[]{ \label{fig:east}
                \includegraphics[trim=2cm 1cm 2cm 0cm,width=\columnwidth]
                {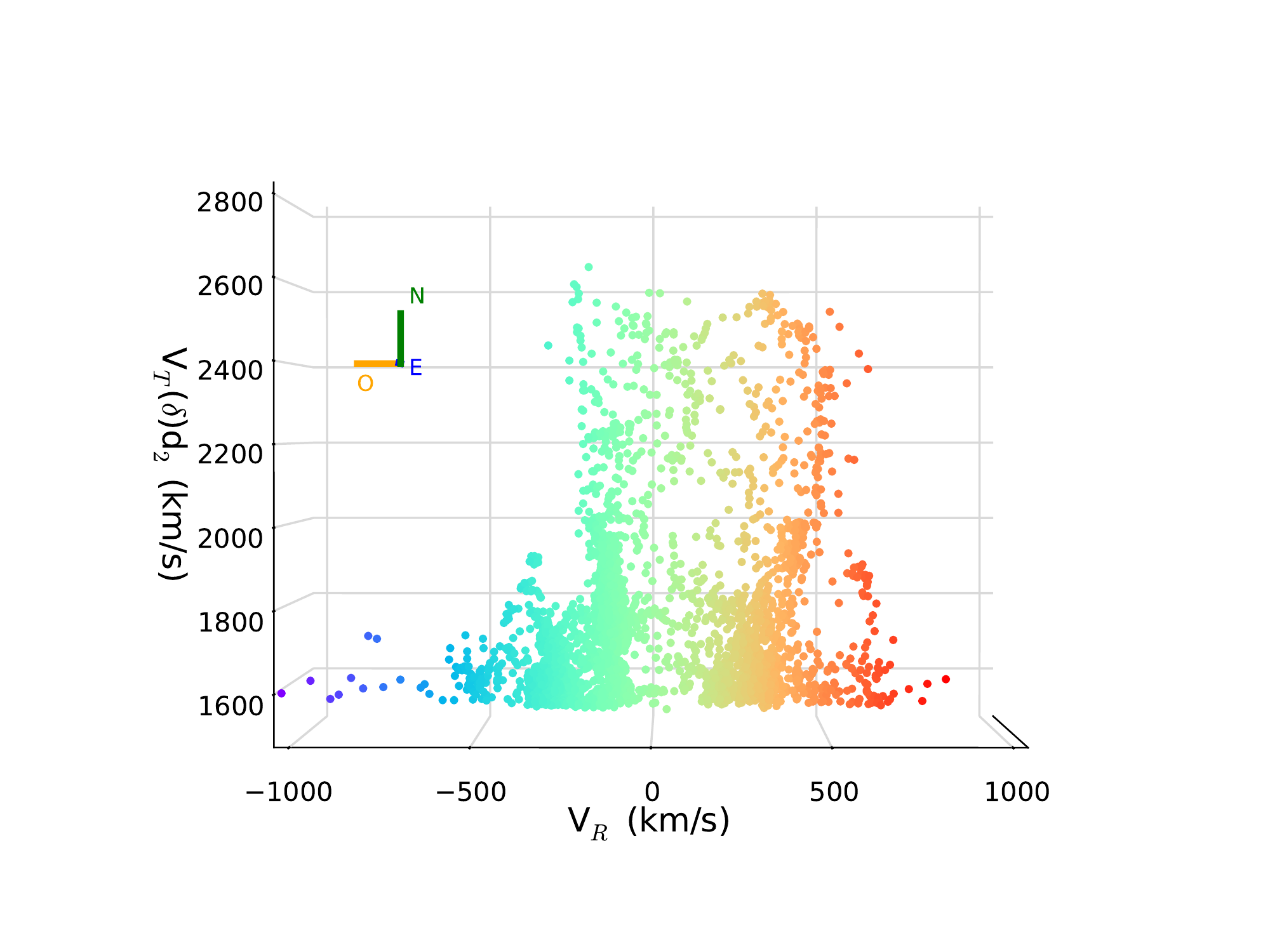} }        
        \subfigure[]{ \label{fig:side1}
                \includegraphics[trim=1cm 0cm .8cm 0.5cm,width=\columnwidth]
                {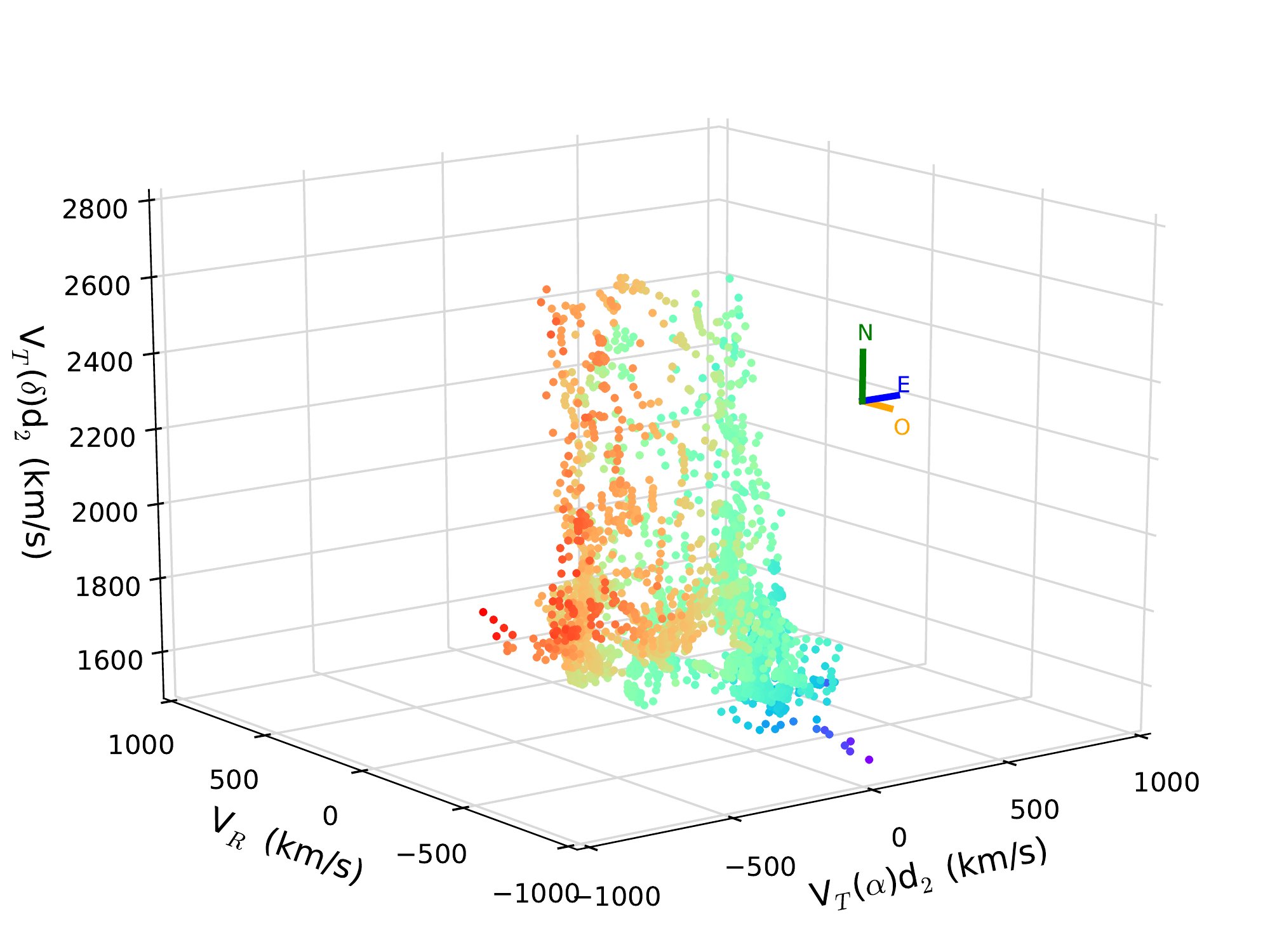} }
%        \subfigure[]{ \label{fig:side2}
%                \includegraphics[trim=.8cm 1cm 1cm 0.5cm,width=\columnwidth]
%                {crab6_3} }
%        \subfigure[]{ \label{fig:west}
%                \includegraphics[trim=2cm 1cm 2cm 0cm,width=\columnwidth]
%                {crab2_3} }

        \caption{3-dimensional views of the Crab Jet.  Red to blue colors correspond to positive and negative radial velocities, respectively. (a) Side view of jet. (b) Angled view from front of jet.  As in Figure \ref{fig:color_sections}, V$_R$ represents radial velocity while V$_T(\alpha)$d$_2$ and V$_T(\delta)$d$_2$ are the transverse velocities in RA and Dec using a distance, d, of 2 kpc. `O' is the direction to the observer, `N' and `E' are the North and East directions on the sky. A color version of this figure is available in the online journal. }
        \label{fig:model}
\end{figure*}

In Figure \ref{fig:rings}, we show the upper seven jet sections when viewed
from four different angles; 0, 45, 90, and 100 degrees.  These plots highlight the jet's slighly elliptical shape, 
its relatively  well defined walls, and the east-west transverse
velocity offset from the remnant's center of expansion.

As the viewing angle is increased, one is increasingly looking down the jet's
major axis. At a viewing angle of 100 degrees, the seven ring sections
appear nearly aligned and the jet is seen as virtually hollow of [\ion{O}{3}]
emission.  We estimate the jet's angle of inclination into the plane of the sky (redshifted) to be
roughly $10 \pm 2$ degrees, a bit more than the previous estimate of eight degrees
by \citet{shu84}.    

\subsection{Radial Velocities}

The average velocities of the most redshifted, most blueshifted, and east and
west sides for the middle six jet ring sections are listed in Table
\ref{tab:ringvel}.  Measured jet section velocities were averaged to minimize bias caused by
our approach of using increased data points to simulate the jet's brighter features.
The maximum blueshifted and redshifted velocities were used to find the
jet's systemic heliocentric velocity which we estimate to be $+170$
$\pm$15 km s$^{-1}$.

%Rad Vel Table
\begin{deluxetable}{lcccc}
  \tablecolumns{5}
  \tablecaption{Radial Velocities of the Jet Ring Sections \label{tab:ringvel}}
  \tablehead{
  \colhead{Jet} & \colhead{Blue side} & \colhead{Red side} & \colhead{East Limb}& \colhead{West 
Limb}\\
  \colhead{Section} & \colhead{(km/s)} & \colhead{(km/s)} & \colhead{(km/s)} & \colhead{(km/s)}\\
    \vspace{-0.3cm}
  }
  \startdata
  	S78 (Maroon) & $-190$ & $+480$ & $+150$ & $+100$ \\
	S65 (Orange) & $-180$ & $+480$ & $+160$ & $+140$ \\
	S50 (Green) & $-170$ & $+475$ & $+175$ & $+195$ \\
	S40 (Cyan) & $-170$ & $+460$ & $(+100)$ & $+195$ \\
	S28 (Blue) & $-160$ & $+440$ & $+170$ & $+200$ \\
	S14 (Purple) & $-145$ & $+415$ & $+170$ & $+205$ \\
   \enddata
   \tablecomments{Radial velocities accurate to $\pm$ 15 km/s}
  \vspace{-0.3cm}
\end{deluxetable}

The expansion velocities listed in Table \ref{tab:ringvel} reveal a few
interesting kinematic properties of the jet.  First, radial velocities of the
jet's eastern limb remain relatively constant with the exception near the top
of the jet where the filaments become more sparse. This velocity continuity suggests
that the jet's eastern limb is fairly uniform despite an appearance of being less
straight and well defined than the jet's western limb.  The radial velocities of the east and west limbs of S40, specifically the east limb, are smaller than expected due to a gap in the ring section which can be seen in Figure \ref{fig:cyan}.

%%% Figure 7: CoE lines
\begin{figure*}
        \centering
        \includegraphics[trim=0cm 0cm 0cm 0cm,width=0.65\textwidth]{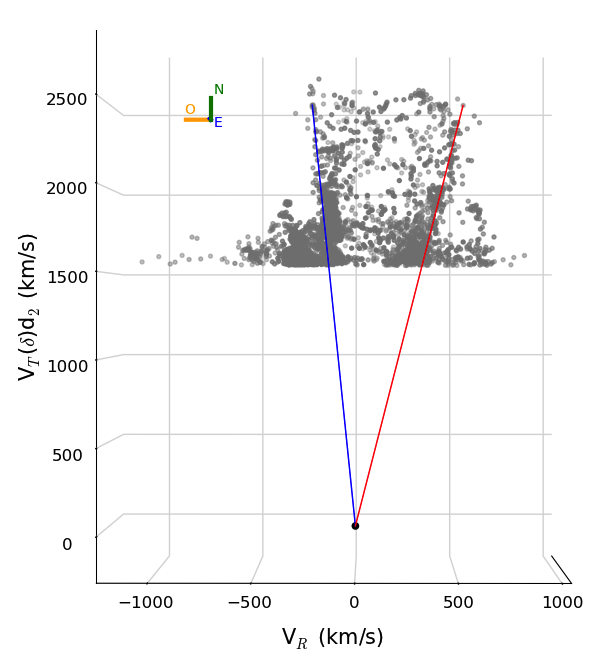}
        \caption{Plot of jet knots and filaments with respect to the Crab Nebula's
expansion center.  Again, V$_R$ and V$_T(\delta)$d$_2$ denote radial and transverse velocities respectively.  A color version of this figure is available in the online journal.}
     \label{fig:coe_lines}
\end{figure*}

In contrast,  radial velocities along the jet's west limb decrease from around
+200 to +100 km s$^{-1}$ from the bottom ring S14 to the upper ring S78
suggesting it is tilted less into the sky plane than the
east limb. 

More significant is the change in radial velocities of the redshifted and
blueshifted sides from the jet's bottom (S14) to the top (S78). The velocities
slowly increase from the base of the jet to its top and a slight increase in
the size of the rings follows this pattern as well.  This increase in radii is
visible in Figure 5 but more easily seen in Figure \ref{fig:color_sections}.
The observed change in velocities over the length of the jet shows the jet
possesses an opening, funnel-like shape rather than being a simple, straight
walled cylinder as its projected parallel edge shape on the sky would indicate.

Figure \ref{fig:model} shows our 3D maps with all emission features plotted,
not just those within our eight jet sections. Red to blue colors correspond to
measured positive to negative radial velocities.  Two viewing angles of the
jet are shown with `O' being the direction of the observer and `N' and `E' are
the North and East directions on the sky. 

From the base to about the middle of the jet, there is a noticeable gradient of
increasing radial velocity.  The top of the redshifted side curves toward lower
velocities, but this should not be mistaken as a cap at the top of the jet.

The plots in Figure \ref{fig:model} show that, with the exception of a handful
of points, the jet's blueshifted side is surprisingly well defined and sharply
edged starting at a N-S transverse velocity (V$_T(\delta)$d$_2$) of 1800 km s$^{-1}$ up to and
including 2600 km s$^{-1}$.  In addition, lower down in the jet from V$_T(\delta)$=1800
to 2000 km s$^{-1}$ a majority of blueshifted jet features are aligned in
parallel with this edge.

In contrast, the jet's redshifted rear side is not as nearly well defined as
its blueshifted side.  Nonetheless, a number of features at the jet's base show
an angled alignment, although not parallel with the upper redshifted jet
regions.    

\subsection{Correlation to The Remnant's Center of Expansion}\label{sec:coe}

It has long been realized that the jet's straight western limb is
roughly aligned back to the Crab's center of expansion
\citep{morRob85,FesGul86,FesSta93}.  This situation 
led us to examine if the front and rear
sides of the jet seen in our 3D maps might also show some 
alignment back to the Crab's center of expansion.

Figure \ref{fig:coe_lines} plots all our measured jet features along with the
Crab's estimated expansion point \citep{Nug98}. The blue and red lines shown
are drawn from the maximum blueshifted and redshifted portions of section S96
back toward the center of expansion. As can be seen, these lines appear in
approximate alignment with the jet's front (blue) and rear (red) sides. 

Of course, there is significant uncertainty inherent in determining such
alignments. The jet is, after all, composed of diffuse filaments and emission
knots and the jet's walls are not perfectly sharply defined. Nonetheless, one
finds good alignment on the facing, blueshifted side but poorer on the rear,
redshifted side. The fact that one sees some alignment at all between the jet's
walls and the projections from the Crab's expansion center, when taken together
with a similar alignment of the jet's western limb, lends support to radially
expanding jet formation models. 

%Figure 8 Spectra and IDL 
\begin{figure*}
         \centering
        \subfigure[]{ \label{fig:crab}
                \includegraphics[trim=0cm 0cm 0cm 0cm,width=.9\columnwidth]
                {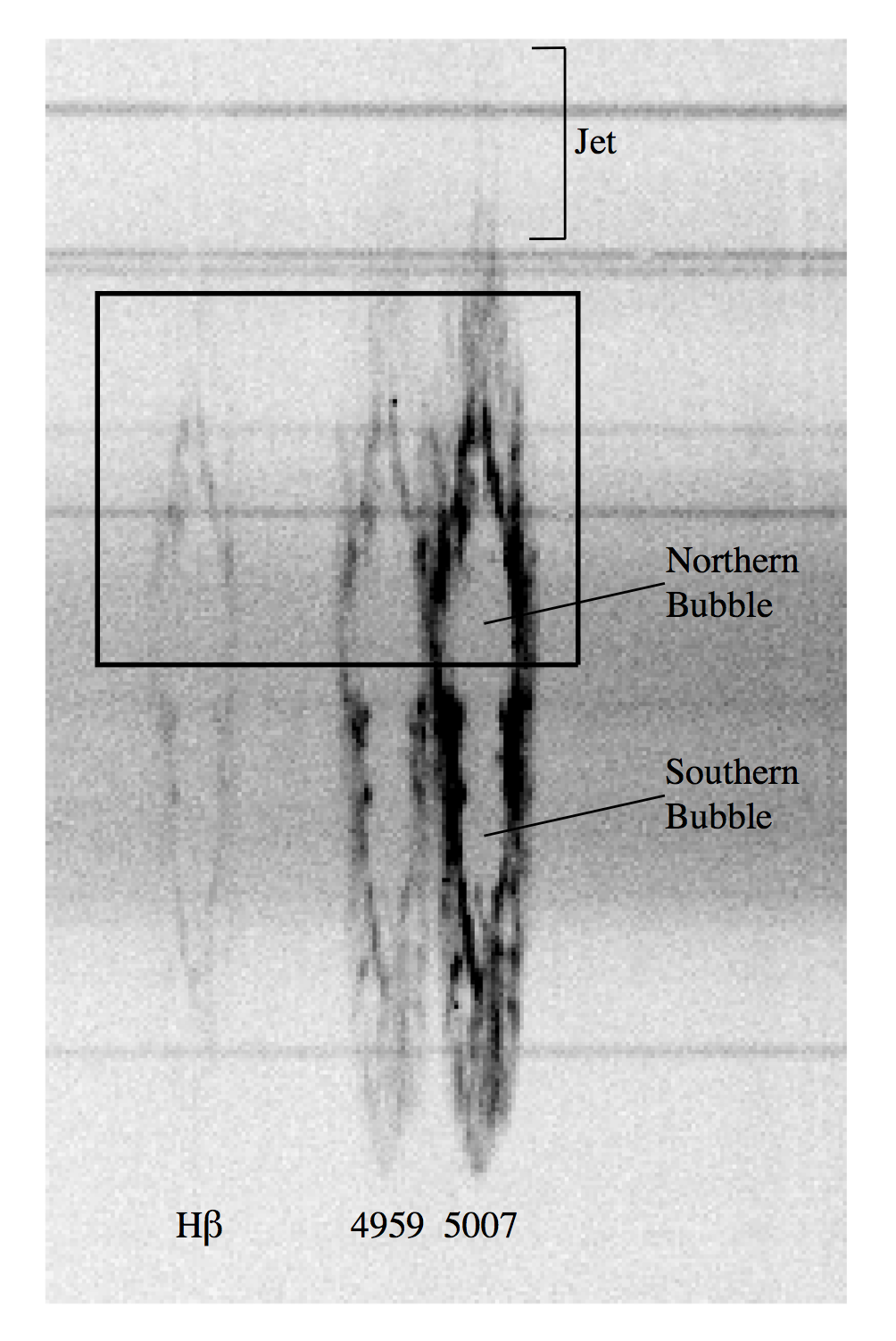} }
                \hfill
        \subfigure[]{ \label{fig:idl2}
                \includegraphics[trim=0cm 0cm 1.3cm 2cm,width=1.1\columnwidth]
                {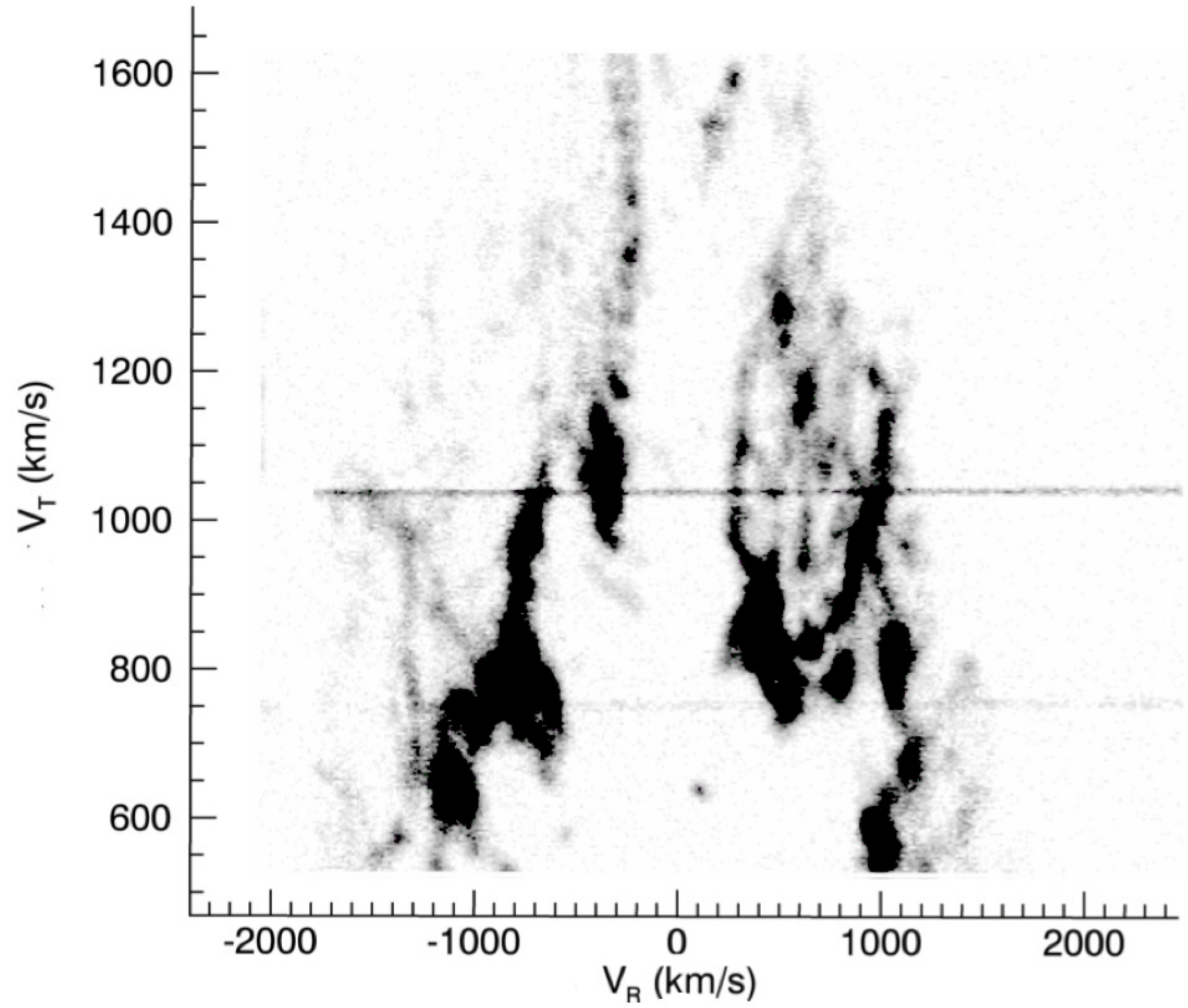} }
        \caption{(a) [\ion{O}{3}] $\lambda\lambda$4959, 5007 and H$\beta$ line emissions
                 from a
                 scanned spectrum of the Crab using a N-S slit \citep{FesShuHur97}.                 
                 (b) Our 3D map of the Crab Nebula from the bottom of the slit to the base of the jet using 
slits 4-8. The box marks the N-S extent of our slit positions used to create the 3D map of the jet-nebula interface.  Note: The dark horizontal lines are stellar continua.}
        \label{fig:base}
\end{figure*}

\subsection{Jet-Nebula Interface}\label{sec:jetneb}

Drifted scanned averaged spectra of the Crab Nebula taken with N-S aligned
slits have long shown that the remnant's expansion is strongly asymmetric,
resembling a N-S bipolar expansion, with the lowest expansion velocities seen
near an east-west belt of exceptional He-rich filaments
\citep{MacA89,FesShuHur97,Smi03,Sat12}.  The jet's location along the Crab's
northern limb places it near the top of this bipolar expansion.

Such low-dispersion scanned spectra suggested a possible opening or channel in the
nebula's thick ejecta shell at the northern extreme of this bipolar structure
and near the base of the jet (see Fig.\ 6 in \citet{FesShuHur97}).
We have used our higher resolution 3D map of the [\ion{O}{3}] emission at the jet's
base and the remnant's northern limb to investigate this possible shell break at the jet/nebula
interface region.

The left hand panel of Figure \ref{fig:base} shows the [\ion{O}{3}]
$\lambda\lambda$4959, 5007 and H$\beta$ line emission profiles from a N-S slit
scanned in the east-west direction across the inner 2 arcmin section of the
Crab obtained by \citet{FesShuHur97}. Like that seen in similar scanned spectra
taken by \citet{MacA89} and \citet{Smi03}, this spectrum shows a systematic velocity
decrease or `pinching' near the remnant's central zone in both [\ion{O}{3}] and
H$\beta$, coincident with the remnant's bright central filaments. Toward the
top of the northern (upper) expansion bubble, the spectra indicate a possible
partial break in an otherwise thick, unbroken shell of emission filaments. In
contrast, no similar break or channel appears present in the southern expansion
bubble. 

The right hand panel of Figure \ref{fig:base} shows our 3D reconstruction of
the [\ion{O}{3}] emission in this region using only slits 4 through 8.  Unlike
the scanned spectra seen in the left hand panel which shows no break in the
northern expansion bubble, our 3D mapping reveals a long and fairly distinct
area nearly devoid of [\ion{O}{3}] emission.  Several emission features
defining this channel can also be seen in the lower resolution scanned
long-slit spectra. The opening in this channel is much clearer in our data and
this difference is presumably due to the wider 2 arcmin coverage of the scanned
long slit data as compared to our 3D mapping which was both more limited in
area and centered on the jet. 

Although largely empty, a few faint filaments lie at the northern end of this
channel and connect with emission seen at the base of the jet and visible in
jet ring section S0 shown in Figure 4a.  The channel's walls connect to the
jet's base at the northernmost limb of the nebula.

\section{Discussion}\label{sec:Disc}

Although it has been over three decades since the jet was first detected and its
peculiar morphology recognized, its origin and nature has remained
uncertain.  Our new spectroscopic data provides a higher resolution view of the
jet's kinematic structure than previous studies, thus offering new insights and
constraints regarding its possible origin.

Despite a striking, parallel tube-like appearance, our 3D mapping shows that
the jet is actually elliptical in shape and wider at the top than its base,
indicative of a funnel-like structure.  This funnel-like structure is not likely the result of lateral expansion due to the synchrotron nebula within the jet because little, if any, synchrotron emission is seen near the top of the jet.
The approximate alignment of the jet's
blue and redshifted sides with the Crab's center of expansion is consistent
with proper motion measurements of individual jet knots and filaments which
show it to be radially expanding with an age matching that of the Crab's
supernova explosion date \citep{FesSta93,RudFesYam08}.

Our spectral reconstruction of the jet's interface with the Crab's thick
filamentary shell revealed a significant hole or `channel' largely free of
[\ion{O}{3}] emission that was only hinted in earlier low dispersion N-S
scanned spectra of the Crab (Fig.\ 8).  The walls of this cavity are closely
aligned with the walls of the jet suggesting a direct connection between the
jet's formation and this opening in the Crab's northern limb.

Taken together, the jet's hollow and funnel-like structure, the radial
alignment of its western limb and blueshifted front side back to the
Crab's center of expansion, an age consistent with the 1054 supernova event,
and its location above a gap in the Crab's emission shell disfavor certain 
proposed theories of the jet's origin.  Its radial motions appear inconsistent
with a RGB mass-loss trail and pulsar beaming models, and models involving
interaction with a local gas cloud near the supernova does not explain the
jet's  lack of emission at its base.

On the other hand, theories of the jet's origin involving the pulsar wind
nebula (PWN) leading to instabilities in the filamentary shell, or a faster
expansion of filaments due to a low density region in the surrounding ISM would
appear consistent with the jet's kinematic and emission properties discussed
above.  The presence of synchrotron emission in the jet \citep{WilWei82,vel84}
lends further support to PWN instability breakout models
\citep{CheGul75,SanHes97,Smi13}.  

However, breakouts of the PWN seen elsewhere along the Crab's outer boundary do
not show any entrained filamentary emission similar to that seen for the
northern jet. Examples of synchrotron emission outside the remnant's sharp
[\ion{O}{3}] emission boundary can be seen along northwest and west limbs of
the Crab in the optical (see our Fig. 1; \citealt{FesMarShu92}, \citealt{Tzi09},
\citealt{Loll13}) and radio \citep{WilSamHog85}.  In addition, PWN instability
breakout models do not make any prediction regarding the jet's northern limb
location especially given its position so far off the NW/SE axis of the
pulsar's wind nebula \citep{Hes02,NgRom04,Kap08}.

In contrast, the transition from the Crab's northern
expansion `bubble' into the channel that leads up to the jet as shown in Figure
8 suggests that if the northern bubble is a result of the initial expansion
post-supernova, the channel and jet are causally related to the supernova event
as well.  This conclusion is consistent with the jet's radial proper motions
and funnel-like shape with walls pointing back towards the Crab's center of
expansion.  A simple explanation for the jet's formation is
that it is the highest velocity component of the Crab's N-S bipolar expansion.

If this picture is correct, it then suggests that the Crab's expansion was
confined to some significant degree in most other directions. The source of
this confinement could be the progenitor's expected super-AGB mass loss wind as
discussed by \citet{Sol01} and \citet{Mor14}.  Based upon the jet's maximum
velocity, this confinement of the Crab's expansion would seem fairly
significant.  Whereas nebula filaments along the northern limb of the Crab
exhibit transverse velocities around 1600 d$_2$ km s$^{-1}$, the location of
the jet's northernmost tip implies a transverse velocities around 2650 d$_2$ km
s$^{-1}$.  This means that if the jet is part of the north bipolar flow, it
extends the northern expansion bubble velocity by 1000 km s$^{-1}$.

Although direct images give the impression the jet arises northward out from
the Crab's northern limb, this is not precisely correct.  The heliocentric
systemic velocity of the Crab is low and effectively 0 km s$^{-1}$ (-20 $\pm$
50 km s$^{-1}$ \citealt{Cla83}, -55 $\pm$ 35 km s$^{-1}$ \citealt{Dav87}, +1 to -7 km
s$^{-1}$ \citealt{Wal99}).  Thus, our estimated +170 km s$^{-1}$ systemic
velocity of the jet means that it is located on the remnant's rear hemisphere,
and tilted some 10 degrees back into the sky plane.

Moreover, the jet's displacement to the east from the center of expansion means
that spectra resulting from east-west scans of the remnant using N-S long slits
are then not strictly aligned with the central axis of the Crab's bipolar
expansion.  Therefore, such spectral scans give an imperfect view of the
remnant's expansion morphology.

In any case, the nature of the remnant's apparent N-S bipolar expansion, of
which the jet lies at the northern tip of, remains unclear.  Although possibly
the signature of a highly asymmetrical supernova explosion, the remnant's
pinched expansion near the east-west band of high helium abundance filaments
\citep{MacA89,law95,Sat12} and the synchrotron's east and west `bays'
\citep{FesMarShu92} may be indictors of circumstellar material that impeded the
expansion \citep{MacA89,FesShuHur97,Smi03,Smi13}.  While the presence of a
pre-SN circumstellar material has also been proposed to help explain both the
early and late optical luminosity of the Crab supernova \citep{Sol01,Smi13}
(but see \citealt{Mor14} for an alternative view), the existence of any pre-SN
circumstellar material around the Crab has not been firmly established
\citep{Mic91,hes95,hes08}. 

\section{Conclusions} \label{sec:Conc} 

The Crab's lone, cylindrical-like `jet' feature along its northern limb has
presented a puzzle ever since its discovery in 1970.  Although faint, its
100$''$ length in a remnant whose radius is less than 200$''$ makes it a
significant feature in the remnant's overall structure and deserves to be
understood.  

Our moderate resolution [\ion{O}{3}] spectra and 3D reconstructions offer the
most detailed look to date at the kinematic and overall structure of the jet
and its interface with the rest of the nebula. From analyses of these data, we
find the following:

1) The jet's systemic velocity is $+170\ \pm$ 15 km s$^{-1}$ with radial
velocities ranging from $-190$ to +480 km s$^{-1}$ and transverse velocities of
1600 to 2650 km s$^{-1}$ from base to tip.  The jet consists of thin
filamentary walls ($V_{exp}$ $\simeq$ 40 -- 75 km s$^{-1}$), is virtually
hollow of [\ion{O}{3}] emission, and elliptical in shape.  Increasing
radial velocities of the red and blue sides of the jet over its length indicate
a funnel-like structure rather than a cylindrical tube as previously thought.

2) The jet's blueshifted facing and redshifted rear sides are relatively well
defined and appear in rough alignment with the remnant's center of expansion.
Such radial alignments suggest a causal connection between the jet's formation
and the Crab's expansion point and do not support jet formation models invoking
a post supernova rupture in the expanding shell of debris along the Crab's
northern limb, or energy flow off the pulsar toward this region. 

3) We find a large and nearly emission-free opening in the remnant's thick
outer ejecta shell located directly below the jet and along the northernmost
point of the Crab's north bipolar expansion. The jet's position at the top of
the bipolar expansion and over this opening strongly suggests the jet to be
part of the Crab's N-S bipolar expansion structure. 

Thus, despite an oddly non-radial and parallel walled appearance, when
considered in the context of the Crab Nebula's bipolar expansion structure, its
origin seems less mysterious.  With an estimated age coincident with the 1054
supernova event, a location atop the remnant's northern expansion bubble
over a large opening in the nebula's ejecta shell, and a radial funnel-like
structure, the jet's origin seems directly linked to the remnant's radial
expansion.  The lack of a similar opening seen in the remnant's ejecta shell in
the Crab's southern expansion bubble may explain the lack of an opposing
jet-like feature along the Crab's southern limb.
 
\acknowledgments
The authors wish to thank Mike Shull and Dan Milisavljevic for providing useful comments and
suggestions, and the staff of MDM Observatory for their
assistance in making these observations possible. This work has made use of
the NASA Astrophysics Data System.  RAF's supernova remnant research is
supported by the National Science Foundation under grant No.\ AST-0908237.

%{\it Facilities:} \facility{MDM}

%\bibliographystyle{apj}
%\bibliography{Crab}

\end{document}